\documentclass{mscs}

\usepackage{graphicx}

\usepackage{amssymb}
\usepackage{amsfonts}
\usepackage{latexsym}
\usepackage[square,comma,numbers,sort&compress]{natbib}

\usepackage{natbib}
\usepackage{makeidx}  % allows for indexgeneration
\begin{document}

\title{The quantum walk search algorithm: Factors affecting efficiency}

\author[Neil B.~Lovett et al.]{Neil B.~Lovett $^{1,2}$, Matthew Everitt $^1$, Robert M. Heath $^1$\thanks{Current address: School of Physics, Heriot-Watt Univeristy, Edinburgh, EH14 4AS, United Kingdom} and Viv Kendon $^1$\\$^1$ School of Physics and Astronomy, University of Leeds, Woodhouse Lane, Leeds, LS2 9JT,\addressbreak United Kingdom,\addressbreak $^2$ Institute for Quantum Information Science, University of Calgary, 2500 University Drive NW,\addressbreak Calgary, Alberta, T2N 1N4, Canada}

%\eads{\mailto{nlovett@ucalgary.ca}, \mailto{V.Kendon@leeds.ac.uk}}

%\pacs{03.67.Ac}

%\noindent{\it Keywords\/}: Quantum walks, Searching

\maketitle

\begin{abstract}
We numerically study the quantum walk search algorithm of Shenvi, Kempe and Whaley [PRA \textbf{67} 052307] and the factors which affect its efficiency in finding an individual state from an unsorted set. Previous work has focused purely on the effects of the dimensionality of the dataset to be searched. Here, we consider the effects of interpolating between dimensions, connectivity of the dataset, and the possibility of disorder in the underlying substrate: all these factors affect the efficiency of the search algorithm. We show that, as well as the strong dependence on the spatial dimension of the structure to be searched, there are also secondary dependencies on the connectivity and symmetry of the lattice, with greater connectivity providing a more efficient algorithm. In addition, we also show that the algorithm can tolerate a non-trivial level of disorder in the underlying substrate.
\end{abstract}

\section{Introduction}

Searching is undoubtedly one of the most basic problems in computer science and computational physics. In this context, searching is not just restricted to a physical database but could also be searching through a state space for an entry which fulfills a specific clause such as the constraint satisfiability problem ($k$-SAT). The classical complexity of such a task scales linearly with the size of the dataset, $N$, to be searched. Intuitively, it is easy to see this must be the case as every item must be checked in turn until the specific item is found. On average, half the items will have to be checked before the correct one is located. This leads to the best classical scaling which can be achieved, $O(N)$.  

One of the most important quantum algorithms discovered thus far is the searching algorithm of Grover \cite{grover96a}. Grover showed that an item could be found from a set of $N$ in a time quadratically faster than the classical case, $O(\sqrt{N})$. Grover's algorithm has been shown to be both optimal and also one of the few quantum algorithms which is provably faster than any possible classical algorithm \cite{bennett97a}. 

Several years after the introduction of this algorithm, Shenvi, Kempe and Whaley \cite{shenvi02a} gave a quantum search algorithm based instead on the discrete time quantum walk, which was first introduced with algorithmic applications in mind by Aharonov et al. \cite{aharonov00a} and Ambainis et al. \cite{ambainis01a}. This quantum walk approach to the search problem is able to match the quadratic speed up of Grover's algorithm. The quantum walk search algorithm has been studied in detail and many improvements have been made since its introduction. In fact, due to the many uses of searching in algorithms, the quantum walk search algorithm has become a standard tool in developing new quantum algorithms \cite{santha08a}. The quantum walk has also recently been shown to be universal for quantum computation and hence a computational primitive \cite{childs09a, lovett10a, underwood10a}, again showing it is a powerful tool.

In \cite{shenvi02a}, the items of the dataset are laid out as the vertices of an undirected graph, specifically a hypercube of dimension $\lceil \log_{2} N \rceil$, on which the quantum walk can be solved analytically \cite{moore01a}. Other recent work by Poto{\v{c}}ek et al.~\cite{potocek08a} has improved the original algorithm by adding an additional coin dimension, allowing the probability of the marked state to approach unity after just one run of the algorithm. This brings the running time of the quantum walk search algorithm very close to the optimal for searching an unsorted dataset, $\pi/4 \sqrt{N}$. Zalka \cite{zalka99a} has previously shown that, for a probability of finding the marked state to be one, this is the best that can be achieved. 

However, the hypercube studied in \cite{shenvi02a} is a highly connected but non-physical structure. In order to make the algorithm more physical, the study of the search algorithm on lower dimensional structures was started by Benioff \cite{benioff00a}. He considered the additional cost of the time it would take a robot searcher to move between different spatially separated data points on $d$-dimensional lattices, stating that in two spatial dimensions, $D$, no speedup was apparent. Subsequently, Aaronson and Ambainis (AA03) \cite{aaronson03a} introduced an algorithm based on a divide and conquer approach, contradicting this claim with a run time of $O(\sqrt{N})$ in dimensions $D\ge3$ and $O(\sqrt{N} \log^{3/2} N)$ when $D=2$.
\begin{table}[t]
\centering
\begin{tabular}{l || c | c | c | c}
Work & $D=2$ & $D=3$ & $D=4$ & $D\ge5$ \\
\hline
AA03 (2003)& $O(\sqrt{N} \log^{3/2} N)$ & $O(\sqrt{N})$ & $O(\sqrt{N})$ & $O(\sqrt{N})$\\
CG03 (2003) & $O(N)$ & $O(N^{5/6})$ & $O(\sqrt{N} \log N)$ & $O(\sqrt{N})$\\
CG04 (2004) & $O(\sqrt{N} \log N)$ & $O(\sqrt{N})$ & $O(\sqrt{N})$ & $O(\sqrt{N})$\\
AKR04 (2004)& $O(\sqrt{N} \log N)$ & $O(\sqrt{N})$ & $O(\sqrt{N})$ & $O(\sqrt{N})$\\
Tulsi (2008) & $O(\sqrt{N \log N})$ & - & - & -\\
Magniez et al. (2008) & $O(\sqrt{N \log N})$ & $O(\sqrt{N})$ & $O(\sqrt{N})$ & $O(\sqrt{N})$\\
Patel et al. (2010) & $O(\sqrt{N \log N})$ & $O(\sqrt{N})$ & $O(\sqrt{N})$ & $O(\sqrt{N})$\\
\end{tabular}
\caption{Summary of runtimes of quantum search algorithms in various dimensions.}
\label{runtime}
\end{table}

Around the same time as this work, Childs and Goldstone (CG03) \cite{childs03a} gave another algorithm, this time based on the continuous time quantum walk, first introduced by Farhi and Gutmann \cite{farhi98a}. They showed a runtime of $O(N)$ for $D=2$, $O(N^{5/6})$ for $D=3$, $O(\sqrt{N} \log N)$ for $D=4$ and $O(\sqrt{N})$ for $D\ge5$. This algorithm is not as efficient as the one introduced in \cite{aaronson03a}, but does represent the first quantum walk search algorithm defined in continuous time. Shortly after this work, Ambainis, Kempe and Rivosh (AKR04)\cite{ambainis04a} gave a discrete time quantum walk search algorithm, improving on the original work of Shenvi, Kempe and Whaley \cite{shenvi02a} by using an additional $\log 2d$ qubits of extra memory. Childs and Goldstone (CG04) \cite{childs04a} later improved their continuous time algorithm by using the Dirac Hamiltonian and hence an additional degree of freedom which can be thought of as adding a coin to the continuous time quantum walk. This approach was able to match that of the discrete time quantum walk search algorithm of Ambainis, Kempe and Rivosh \cite{ambainis04a}. These results are summarised in table \ref{runtime}. Up to this point, it remained an important open question as to whether the full quadratic speedup could be achieved in two spatial dimensions. 

It took several years for any further improvements to be found in two spatial dimensions. Tulsi \cite{tulsi08a} then managed to improve the run time for $D=2$ by a $\sqrt{\log N}$ factor to $O(\sqrt{N \log N})$ using a modified version of the algorithm with ancilla qubits. In the previous cases, the probability of the marked state scaled logarithmically with the size of the data set, $O(1/\log_{2} N)$. In his work, Tulsi is able to control this probability using the ancilla qubits to give a constant scaling of the probability at the marked state, $O(1)$, thus removing the need for the $\sqrt{\log N}$ amplitude amplification steps. During the years prior to the work by Tulsi, several advances were made in establishing a theory of quantum walk search algorithms. This was pioneered by Szegedy \cite{szegedy04a} who was able to introduce a method to quantise classical Markov chains (classical random walks on graphs) based on the previous work of Ambainis \cite{ambainis04c}. This framework is similar to other work by Ambainis, Kempe and Rivosh \cite{ambainis04a} and both have been used to develop algorithms which give complexity gains compared to the basic Grover search \cite{magniez05a,magniez05b,buhrman04a}. Building on all these approaches, Magniez et al. \cite{magniez07a} developed a quantum walk search algorithm for any quantum walk based on a reversible, ergodic (a stationary distribution can be found) classical Markov chain. This extends previous work as the algorithm is applicable to a much larger class of Markov chains. It also combines previous ideas into one coherent theory of quantum walk search algorithms. Following this work, Magniez et al. \cite{magniez08a} gave a similar theory for the hitting times of quantum walks. They prove that, given a reversible, ergodic classical random walk, the hitting time of the equivalent quantum walk is quadratically faster than the classical case. In addition, they actually prove this speedup is tight for a large class of these quantum walks where the unitary operation is a reflection. It is well known that the hitting time of a classical random walk on a 2D lattice is $O(N \log N)$. Therefore, the equivalent quantum walk hitting time would be $O(\sqrt{N \log N})$ which then matches the run time of Tulsi \cite{tulsi08a}. Magniez et al. \cite{magniez08a} also show they can find the probability of the marked state in a constant fashion, thus extending the result of Tulsi \cite{tulsi08a} to the larger class of quantum walks which are based on reversible, ergodic Markov chains. In fact, this result has recently been tightened further by Krovi et al. \cite{krovi10a}, showing that the classical Markov chain, which forms the basis of the quantum walk, need only be reversible. These results tend to indicate that it is unlikely the additional $\sqrt{\log N}$ factor in the run time of the algorithm in two spatial dimensions can be removed. Other recent work shows similar results, including Marquezino et al. \cite{marquezino10a} who show the mixing time of a quantum walk on a two dimensional toroidal lattice is also $O(\sqrt{N \log N})$. Additionally, a different approach to the searching problem, using the staggered lattice fermion formalism, has been put forward by Patel et al. \cite{patel10a,patel10b} to give the same run time. In related work, Hein and Tanner \cite{hein10a} give a detailed analysis of the search algorithm on $d$-dimensional lattices in terms of the level dynamics near an avoided crossing. They find the same additional $\sqrt{\log N}$ factor in the run time of the algorithm in two spatial dimensions. They also give analytical expressions for the prefactors to the basic scaling of both the time to find the marked state and also the maximum probability the marked state reaches. All of these results lend further weight that the two dimensional case is the critical dimension and it is unlikely that the full quadratic speedup is possible.

Almost all previous studies of the quantum walk search algorithm have focused on the dependence the algorithm has on the spatial dimension of the structure being searched. Little has been done to explore other factors which may affect the runtime and hence the efficiency of the algorithm. This is due to the connectivity or lack of symmetry within interesting structures making them hard to analyse analytically. However, Abal et al. \cite{abal10a} have shown analytically that the complexity of the search algorithm on the hexagonal lattice is $O(\sqrt{N} \log N)$, matching the search on the Cartesian lattice in \cite{ambainis04b} but with a differing prefactor to the scaling of the algorithm. In addition, highly symmetric graphs such as the complete graph were studied by Reitzner et al. \cite{reitzner09a} showing the additional connectivity does not allow the search to beat the optimal lower bound of $O(\sqrt{N})$. The hitting time on the complete graph has also been studied recently by Santos and Portugal \cite{santos09a}, proving this is also $O(\sqrt{N})$. Finally, the constant prefactors to the $O(\sqrt{N})$ scaling on the hypercube and $d$-dimensional lattices have been determined analytically in work by Hein and Tanner \cite{hein09a, hein10a}.

In this work, we investigate numerically the factors which affect the efficiency of the search algorithm in terms of the prefactors to the scaling of both the maximum probability of the marked state and also the time to find this maximum probability. After describing the discrete time quantum walk and how it can easily be modified to become a search algorithm in the next section, we move on to investigate how the quantum walk search algorithm is affected by the dimensionality of the underlying substrate, sec.~\ref{sec:dimensionality}. We introduce a simple form of tunnelling to allow us to interpolate between structures of differing spatial dimension. In sec.~\ref{sec:connectivity}, we move on to study how varying the connectivity of regular structures impacts the prefactors to the scaling of the algorithm. This also uses the same form of tunnelling and extends previous work by Lovett et al. \cite{lovett10b}. The final factor we investigate, sec.~\ref{sec:disorder}, is disorder in the underlying substrate. We model this using percolation lattices in both two and three dimensions to establish how much, if any, disorder the search algorithm can tolerate whilst still maintaining a quantum speed up. Finally, we conclude with a discussion of our findings in sec.~\ref{sec:discussion}.
 
%%%%%%%%%%%%%%%%%%%%%%%%%%%%%%%%%%%%%%%%%
\section{The quantum walk search algorithm}
\label{sec:searchalgorithm}
%%%%%%%%%%%%%%%%%%%%%%%%%%%%%%%%%%%%%%%%%

\subsection{Discrete time quantum walk on the infinite line}

A discrete time quantum walk on the line is defined in direct analogy with
a classical random walk. In the quantum case, the walker is replaced by a quantum particle carrying a two state quantum system for the coin. In order to maintain quantum dynamics, which must be reversible, the `coin toss' is effected by a unitary operator. We denote the basis states for the quantum walk as an ordered pair of labels in a `ket' $|x,c\rangle$, where $x$ is the position and $c\in \{0, 1\}$ is the state of the coin. The walker is started at the origin with an internal coin state of 0. At each timestep we act on the quantum walker with a coin operator followed by a conditional shift operator.

The simplest coin operator is the Hadamard $H$, defined by its action
on the basis states as
\begin{eqnarray}
H | x, 0 \rangle &= \frac{1}{\sqrt{2}}(| x, 0 \rangle + | x, 1 \rangle) \nonumber \\
H | x, 1 \rangle &= \frac{1}{\sqrt{2}}(| x, 0 \rangle - | x, 1 \rangle),
\label{hadamard}
\end{eqnarray}
and the shift operation $S$ acts on the basis states thus
\begin{eqnarray}
S | x, 0 \rangle &= | x-1, 0 \rangle \nonumber \\
S | x, 1 \rangle &= | x+1, 1 \rangle.
\label{shift}
\end{eqnarray}
The coin operator splits the walker into a superposition of coin states and the conditional shift operator then moves the walker to the correct position based on the coin state. The first three steps of a discrete time quantum walk starting from the origin, in coin state 0, are\\
\begin{eqnarray}
(SH)^3| 0, 0 \rangle &=&  (SH)^2 S \frac{1}{\sqrt{2}}(| 0, 0 \rangle \ + | 0, 1 \rangle) \nonumber \\
&=& (SH)^2 \frac{1}{\sqrt{2}}(| -1, 0 \rangle \ + | 1, 1 \rangle) \nonumber \\
&=& (SH) S \frac{1}{2}(| -1, 0 \rangle \ + | -1, 1 \rangle \ + | 1, 0 \rangle \ - | 1, 1 \rangle) \nonumber \\
&=&  SH \frac{1}{2}(| -2, 0 \rangle \ + | 0, 1 \rangle \ + | 0, 0 \rangle \ - | 2, 1\rangle) \nonumber \\
&=& S \frac{1}{\sqrt{8}}(| -2, 0 \rangle \ + | -2, 1 \rangle \ + | 0, 0 \rangle \ - | 0, 1 \rangle \ + | 0, 0 \rangle \ + | 0, 1 \rangle \nonumber \\
 \ &-& | 2, 0 \rangle \ + | 2, 1 \rangle) \nonumber \\
&=& \frac{1}{\sqrt{8}}(| -3, 0 \rangle \ + | -1, 1 \rangle \ + 2 | -1, 0 \rangle \ - | 1, 0 \rangle \ + | 3, 1 \rangle).
\label{threesteps}
\end{eqnarray}

As the walk progresses, quantum interference occurs whenever there is more
than one possible path of $t$ steps to the position.
This can be both constructive and destructive, as shown in eq.~(\ref{threesteps}),
which causes some probabilities to be amplified or decreased at
each timestep. This leads to the different behaviour compared to its classical counterpart: spreading at a rate proportional to $t$, quadratically faster than the classical random walk. In addition, the centre part of the distribution, in the interval $[ -t/\sqrt{2}, t/\sqrt{2}]$, is fairly uniform. This is the opposite of the classical random walk which has an exponential drop in probability after just a few standard deviations from the origin. These properties of the quantum walk on the line were obtained by both Ambainis et al. \cite{ambainis01a} and Nayak and Vishwanath \cite{nayak00a}. 

\begin{figure}[!tb]
\begin{minipage}{\columnwidth}
    \centering
	\includegraphics[width=0.85\columnwidth]{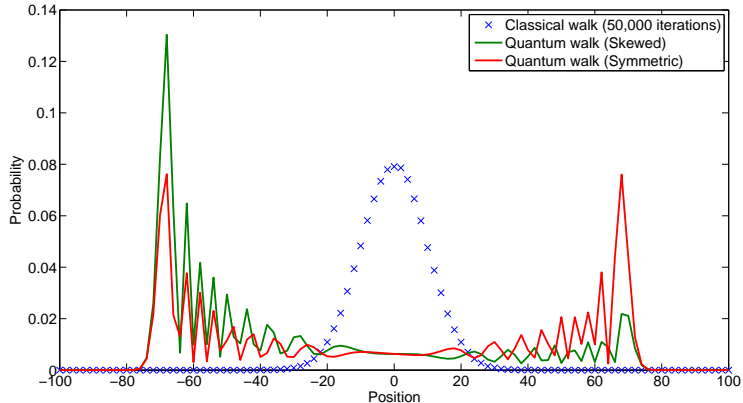}
	\caption[Probability distributions for a classical and quantum walk.]{Classical (crosses) and quantum (solid lines) probability
	distributions for walks on a line after 100 timesteps.
	Only even positions are shown since odd positions are zero. The classical walk is averaged over 50,000 iterations of the random walk. A skewed quantum walk is shown with an initial state of $| 0,0\rangle$ along with a symmetric quantum walk with an initial state of either $ \sqrt{0.15}| 0,0\rangle + \sqrt{0.85}| 0,1\rangle$ or $1/\sqrt{2} (| 0,0 \rangle + i| 0,1\rangle)$.}
    \label{qandcwalk}
\end{minipage}
\end{figure}

As the walker can now be in a superposition of positions on the line, we obtain a probability distribution of the quantum walker after one run of the entire walk. Obviously, this is due to the fact the coin operator is now deterministic. However, if we were to measure the coin after the required number of timesteps, we would get a random output as in the classical case. We show both the classical and quantum probability distributions after 100 steps in fig.~\ref{qandcwalk}. If the walk is imperfect and some decoherence is allowed, we can see the gradual change from the quantum case back to classical. Kendon et al. \cite{kendon03c} investigated this in detail showing that as the decoherence in the system grows, the spread of the walk gradually changes from the quantum walk shown above back to the classical binomial distribution. In the interim, we see a gradual change with an almost `top-hat' distribution being found which is useful for random sampling. For a review of the effects of decoherence in quantum walks, see Kendon \cite{kendon06a}. 
%------------------%
%------------------%

\subsection{Discrete time quantum walk in higher dimensions}

We can see that the quantum walk exhibits interesting and very different behaviour to the classical walk even on the line. However, many interesting problems in computer science are defined in higher dimensions. In order to define the walk in these higher dimensional, we require a new coin operator, of dimension $d$, in order to span the entire coin state space of the walker \cite{moore01a, mackay02a, kendon03b, kempe03a}. This can be any unitary operator of the required dimension. Clearly many different possibilities exist but we only mention the most common operator here - the Grover coin,
\begin{equation}
G^{(d)} = \left( \begin{array}{ccc}  \frac{2}{d} & \dots &  \frac{2}{d} \\  \vdots & \ddots & \vdots \\  \frac{2}{d} & \dots &  \frac{2}{d} \end{array} \right ) - I_{d},
\label{grovercoin}
\end{equation}
where $d$ is the degree of the vertex and $I_{d}$ is the identity operator of the same dimension. The Grover coin is symmetric but only balanced, i.e. it treats all directions in the same way - up to a phase factor, for the cases where $d=2$ and $d=4$. In the case of $d=3$ and all higher dimensions, the coin treats one edge differently to the remaining $d-1$. In addition to the coin operator, the conditional shift operator must also be modified. In the case of the line, it is easy to define as there are only two possible directions the walker can move in. In higher dimensions, the walker can move in any one of $d$ directions. Kendon \cite{kendon03b} treats this problem rigorously, but the most important thing is to maintain a consistent labelling approach for each of the edges. 

\begin{figure}[!tb]
\begin{minipage}{\columnwidth}
    \centering
	\includegraphics[width=0.49\columnwidth]{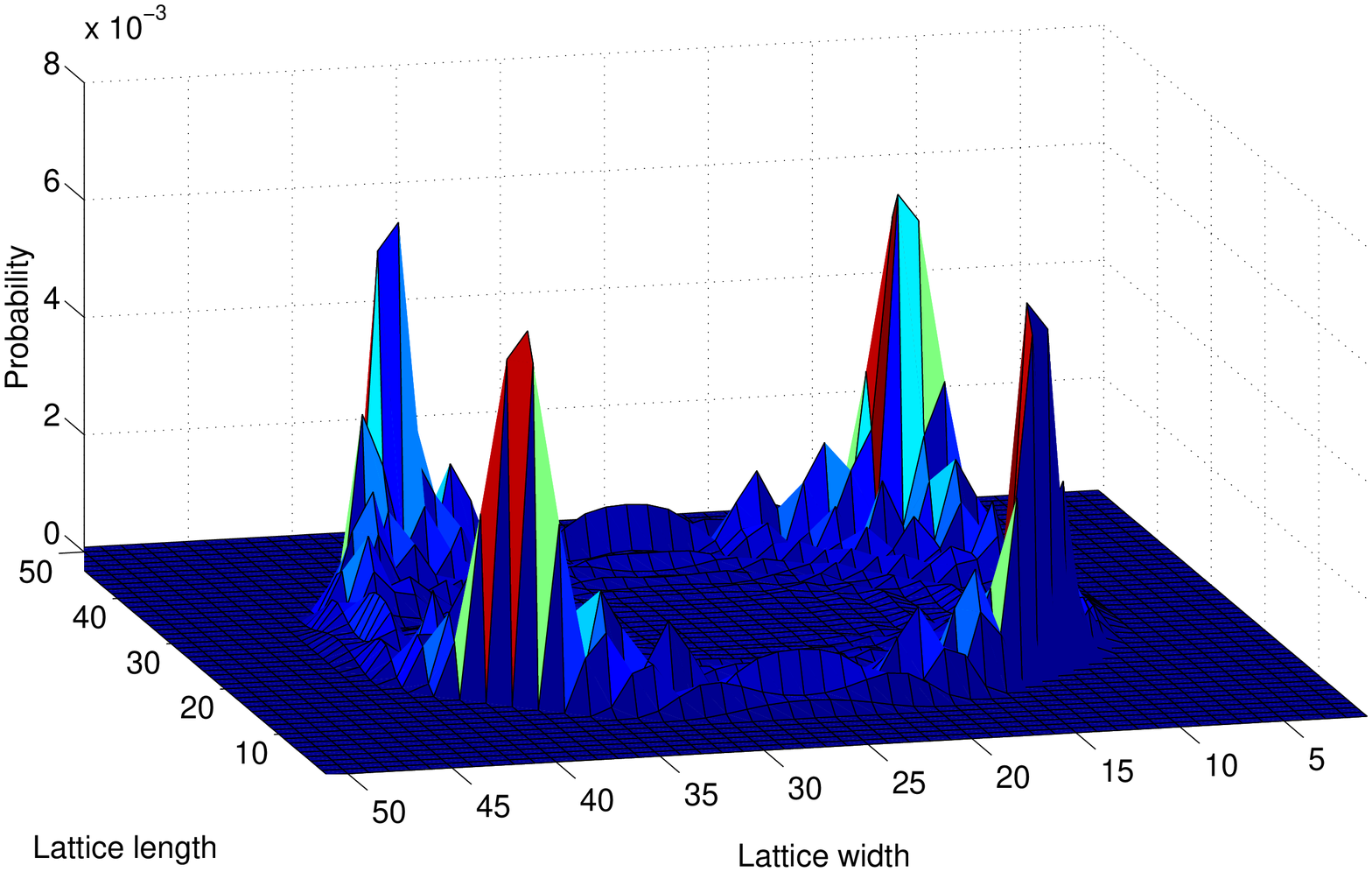}
	\includegraphics[width=0.49\columnwidth]{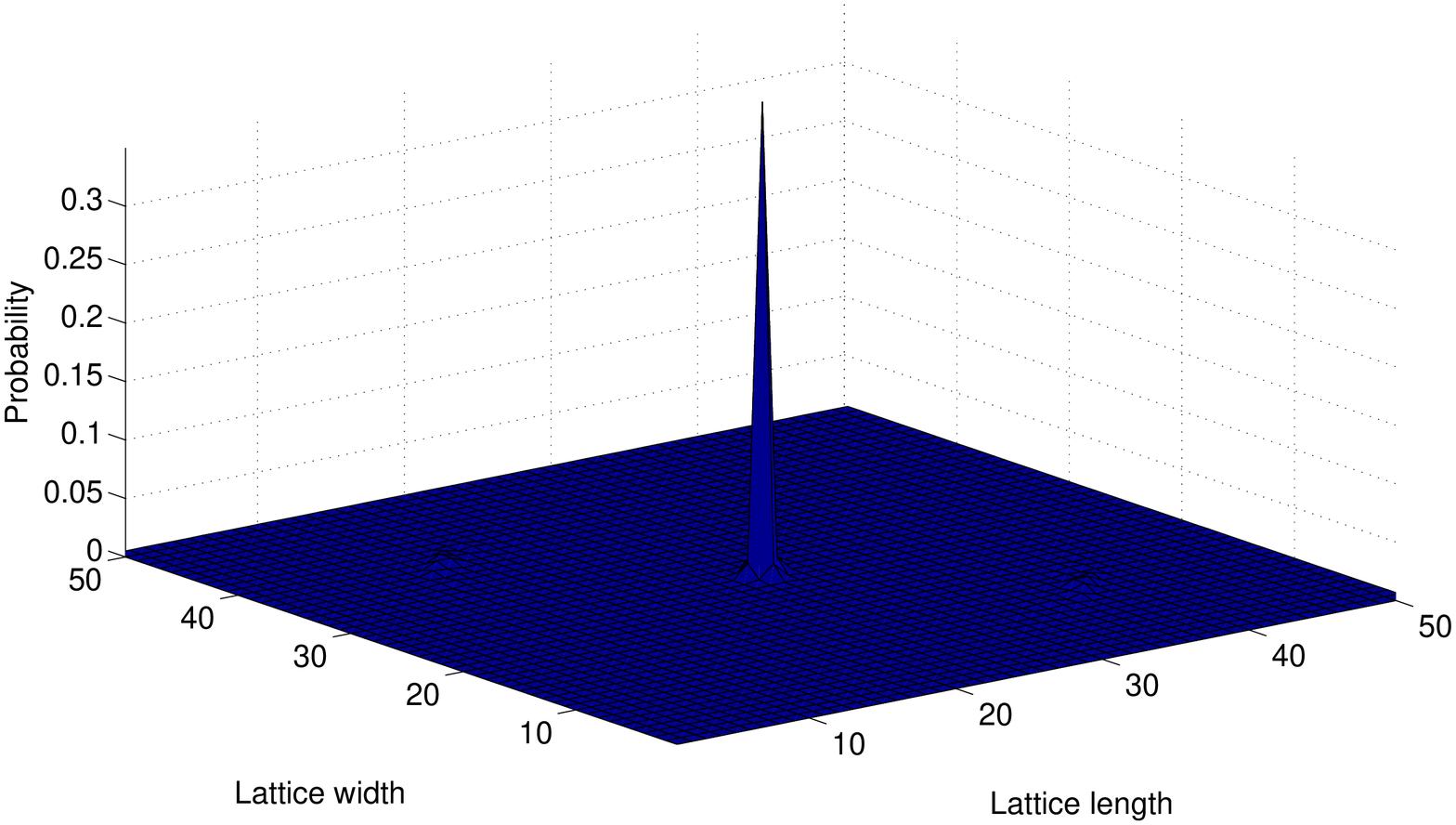}
	\caption[Dynamics of the quantum walk on a two dimensional Cartesian lattice using the Grover coin.]{Dynamics of the quantum walk on a two dimensional Cartesian lattice using the Grover coin, eq.~(\ref{2dgrovercoin}). LHS: Maximum spreading obtained using the initial state in eq.~(\ref{grovermaxinitialstate}). RHS: Localisation obtained using the initial state in eq.~(\ref{syminitialstate}). Note the different scales.}
    \label{grovercoindynamics}
\end{minipage}
\end{figure}

The dynamics of the walk on higher dimensional structures has been studied briefly by Mackay et al. \cite{mackay02a} and then in more detail by Tregenna et al. \cite{kendon03a}. They numerically studied the spreading of the quantum walk with varying coin operators and initial states. Tregenna et al. found that the initial state of the walker on the lattice had a large impact on the spreading of the walker. Depending on the initial state, the walker can spread anywhere from a minimum possible spread to a maximum possible spreading (as defined in \cite{kendon03a} by the second moment). Using the Grover coin on a two dimensional Cartesian lattice, $d=4$, 
\begin{equation}
G^{(4)} = \frac{1}{2} \left ( \begin{array}{cccc} -1 & 1 & 1 & 1 \\ 1 & -1 & 1 & 1 \\ 1 & 1 & -1 & 1 \\ 1 & 1 & 1 & -1 \end{array} \right ),
\label{2dgrovercoin}
\end{equation}
most of the initial states, including the symmetric initial state,
\begin{equation}
| \psi\rangle_{sym} = \frac{1}{2} \left ( | 0,L \rangle + i| 0,R \rangle + |0,D\rangle + i|0,U\rangle \right ),
\label{syminitialstate}
\end{equation}
where $L, R, D, U$ are the four directions the walker is able to move on the lattice, give a minimal spreading with the walker localising around the origin with high probability as seen in fig.~\ref{grovercoindynamics}. However, one specific state, 
\begin{equation}
| \psi\rangle_{max} = \frac{1}{2} \left ( | 0,L \rangle - | 0,R \rangle + |0,D\rangle -|0,U\rangle \right ),
\label{grovermaxinitialstate}
\end{equation}
gives a maximal spreading, again shown in fig.~\ref{grovercoindynamics}. Inui et al. \cite{inui04a} proved this localisation analytically for two dimensional lattices. In addition to these results, some analytical results have been shown for $d$-dimensional lattices. Grimmett et al. \cite{grimmett04a} proved that in the limit $n \to \infty$, $X_{n}/n$ converges weakly where $X_{n}$ is the position at time $n$ in the case of the infinite line. Gottlieb et al. \cite{gottlieb05a} later extended this result to show convergence on $d$-dimensional lattices.

\subsection{The discrete time quantum walk search algorithm}

\begin{figure}[!tb]
    \centering
	\includegraphics[width=0.45\textwidth]{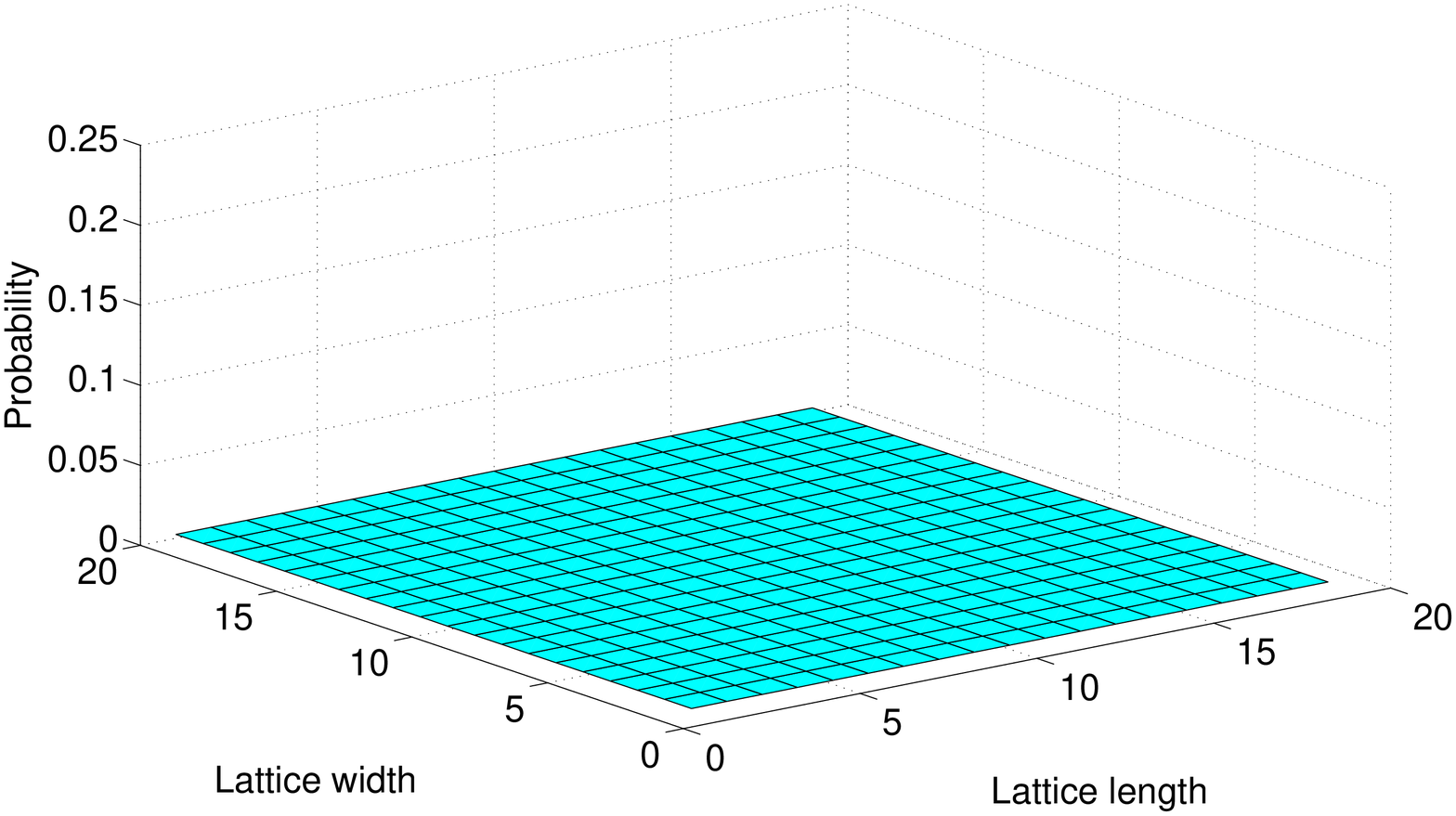}
	\includegraphics[width=0.45\textwidth]{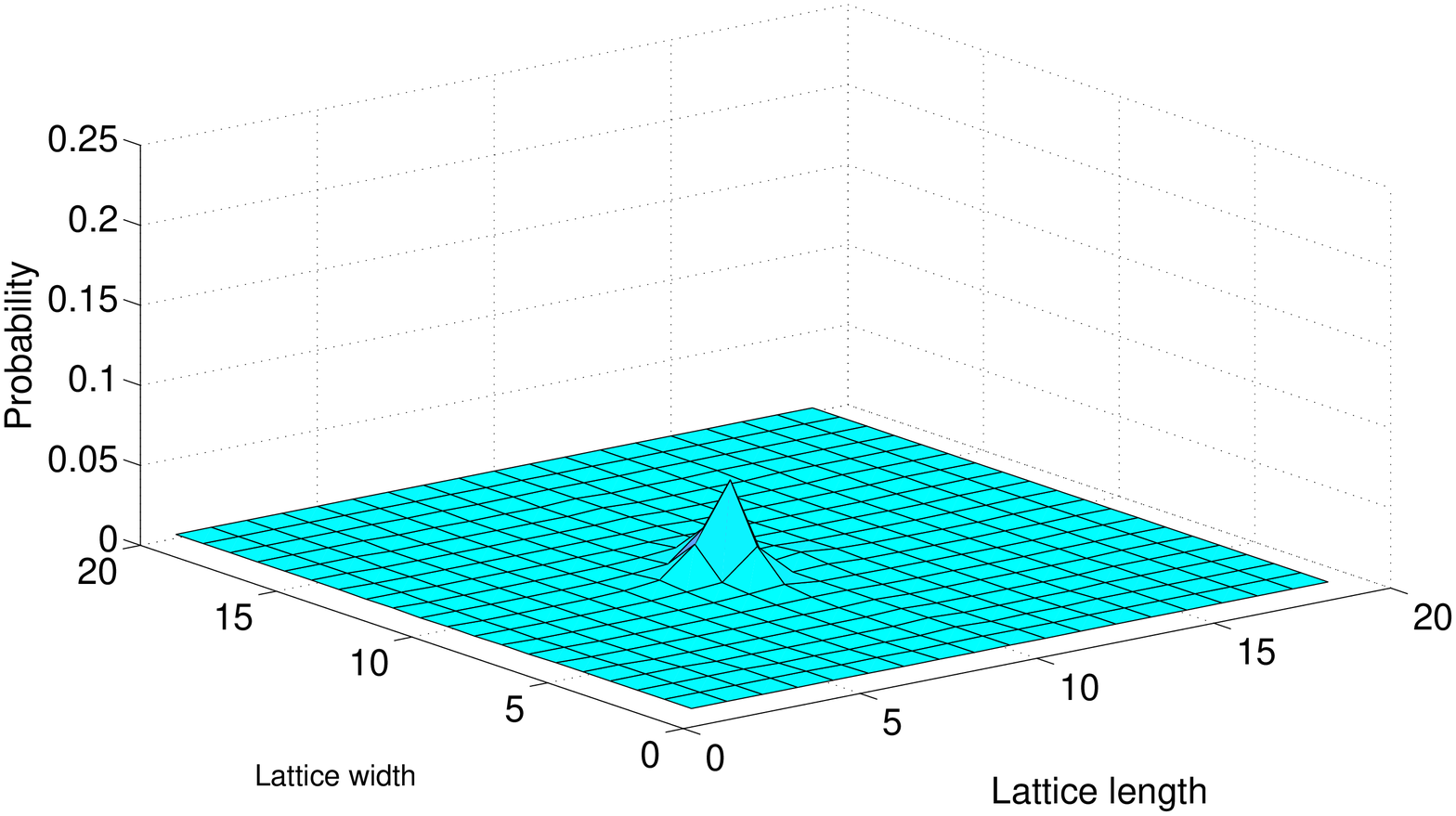}
	\includegraphics[width=0.45\textwidth]{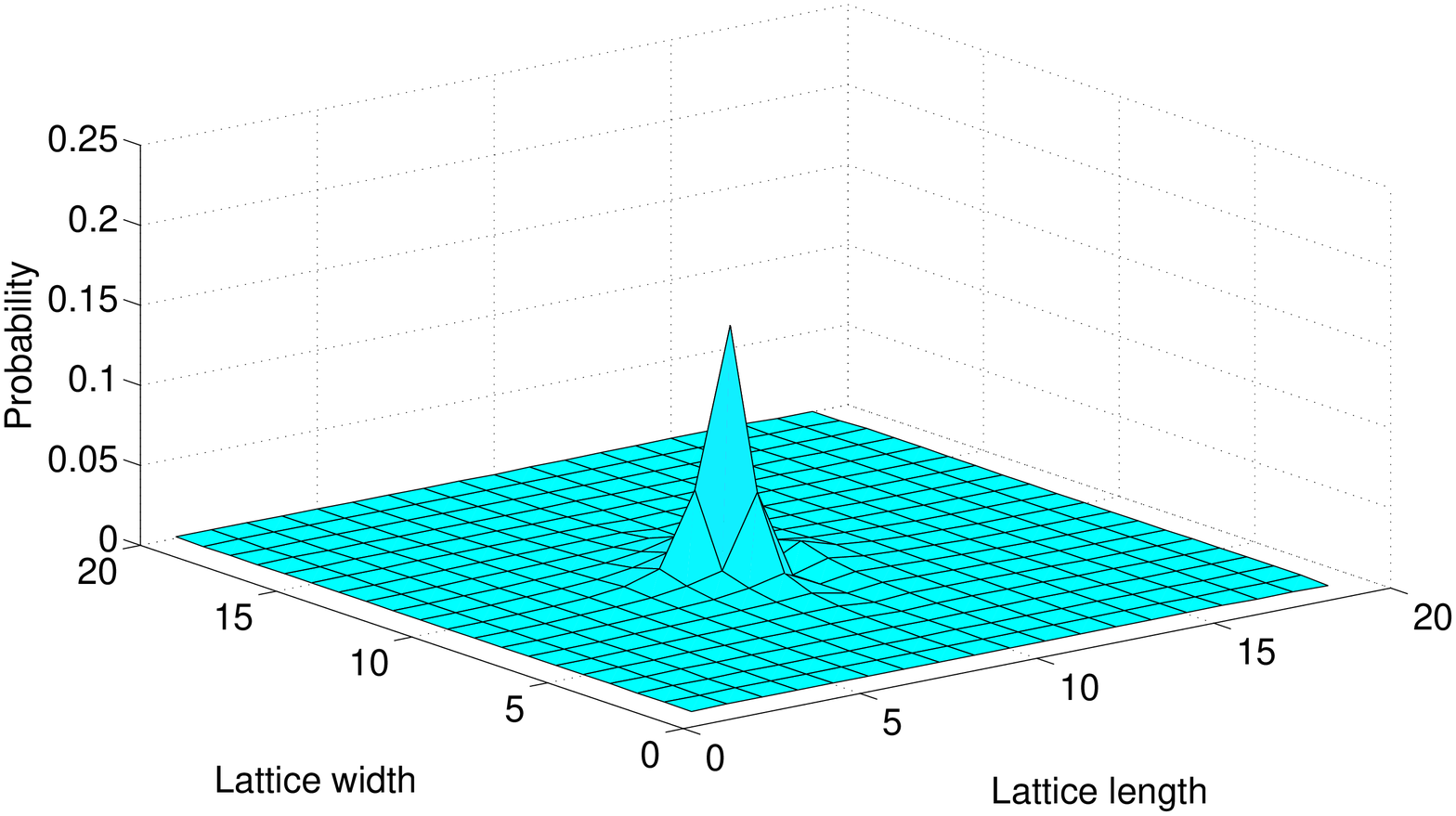}
	\includegraphics[width=0.45\textwidth]{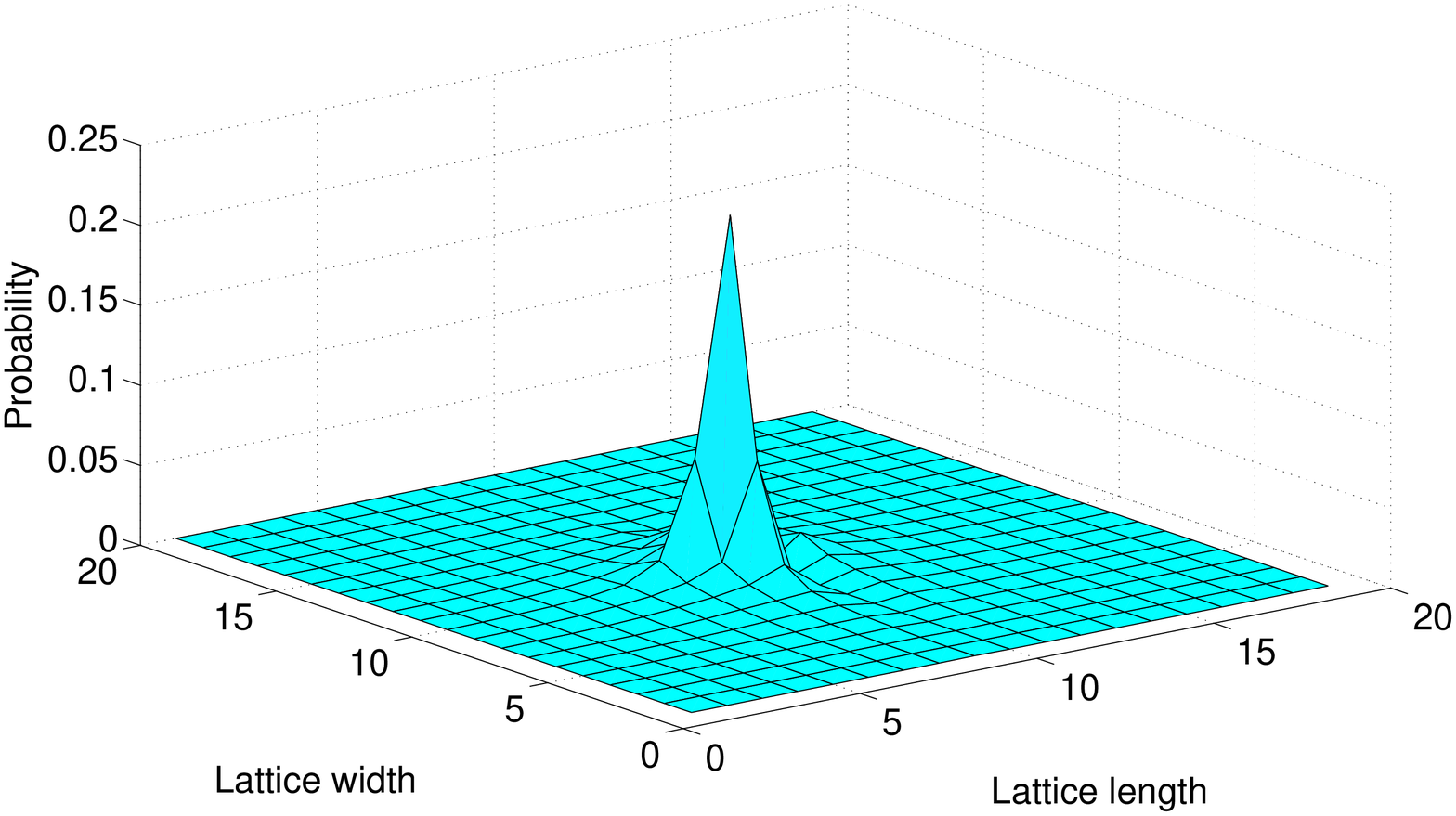}
	\caption[Probability distributions of a discrete time quantum walk
	search at various timesteps.]{Probability distribution of a discrete time quantum walk
	search on 400 vertices arranged in a $20\times 20$ square
	with periodic boundary conditions, evolved for 0, 10, 20 and 32
	timesteps. The marked vertex is at position 190.}
    \label{twod4set}
\end{figure}

\begin{figure}[!tb]
    \begin{minipage}{\columnwidth}
	\centering
	\includegraphics[width=0.7\textwidth]{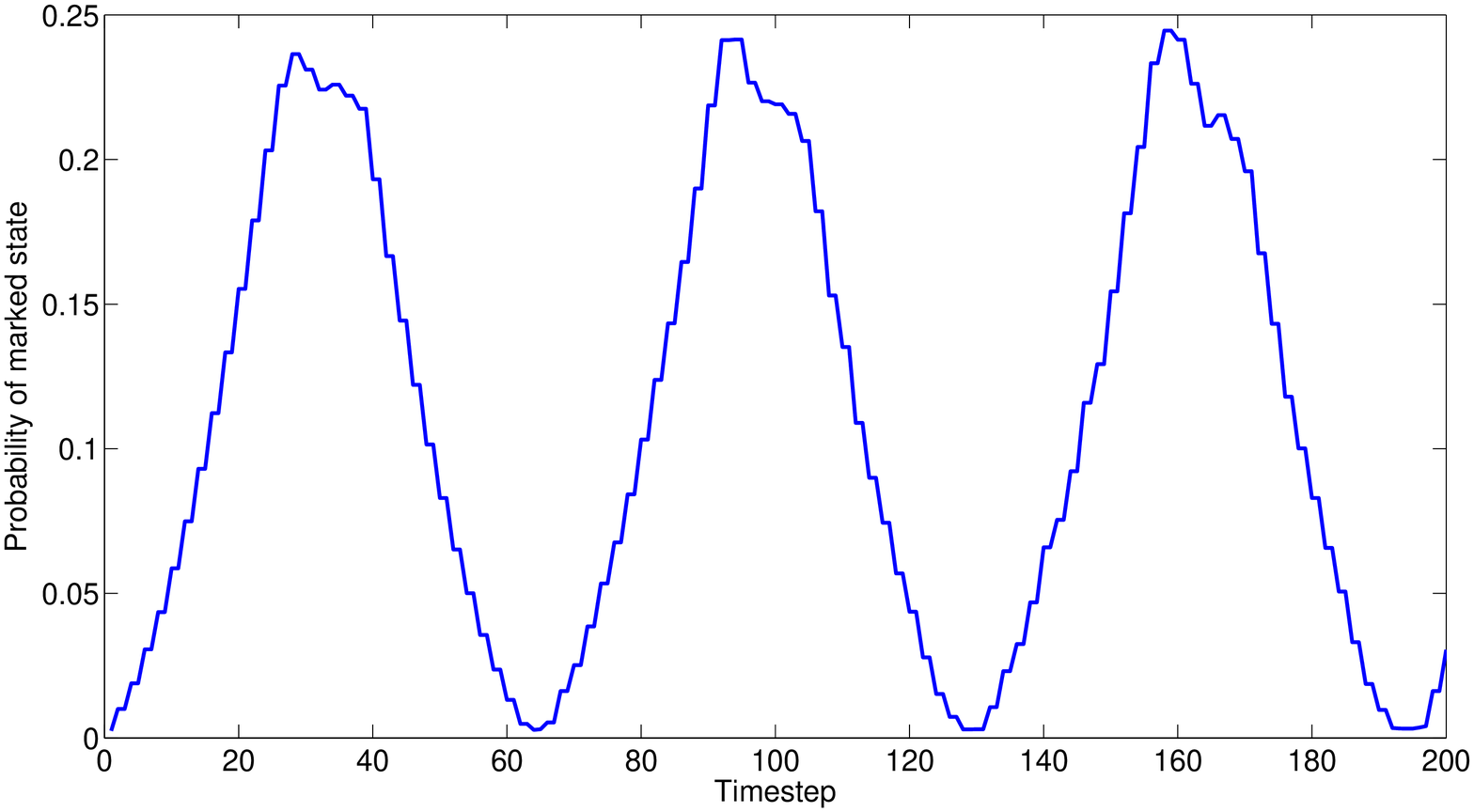}
	\caption[Probability distribution of the marked state over time.]{Probability of the marked state over 200 timesteps
	on a $20\times 20$ grid with periodic boundary conditions.
	The marked vertex is at position 190.}
	\label{maxprob2d}
    \end{minipage}
\end{figure}

We now describe how Shenvi, Kempe and Whaley \cite{shenvi02a} were able to modify the quantum walk into a search algorithm. In their work, they analysed the search algorithm on a hypercube. Here, we show how the quantum walk search algorithm is applied to a 2D Cartesian lattice. The data points we wish to search are laid out as the vertices of an undirected graph. The edges then represent the specific connections between data points. At the edges of the lattice we impose periodic boundary conditions, in effect turning the graph into a torus. Our aim is to find one data item, a specific vertex, out of the set of data to be searched. We start the walker in an equal superposition of all the possible sites in the lattice, and the coin in an equal superposition of all directions,  
\begin{equation}
|\psi\rangle = \frac{1}{\sqrt{dN}}\sum_{x=1}^N\sum_{c=1}^d |x,c\rangle,
\end{equation}
where $d$ is the degree of the vertices in the graph and $N$ is the total number of vertices. If we let the walker evolve in a natural fashion, using the Grover coin eq.~(\ref{grovercoin}), we would find a flat distribution identical to the starting state at any point in time. This uniform distribution is an eigenstate of the Grover coin operator. We need to use a different coin operator for the marked state in order to introduce a bias into the walk. It is optimal to invert the phase of the $G^{(4)}$ coin operator from eq.~(\ref{2dgrovercoin}), as shown in \cite{lovett10b}, giving
\begin{equation}
G^{(4)}_m = \frac{1}{2}\left( \begin{array}{cccc} 1 & -1 & -1 & -1 \\ -1 & 1 & -1 & -1 \\ -1 & -1 & 1 & -1 \\ -1 & -1 & -1 & 1 \end{array} \right ).
\label{markedcoin2d}
\end{equation}
Figure~\ref{twod4set} shows how the distribution of the
walker evolves with time for a $20\times 20$ lattice, i.e. $N=400$. We see that using a different coin creates a defect in the walk and the probability coalesces on the marked state over time.
%------------------%
%------------------%
As the walk progresses, the probability at the marked state cannot keep increasing without limit. In fact, we see in fig.~\ref{maxprob2d} that the probability at the marked state has periodic behaviour with the first peak occuring at roughly $t=(\pi/2)\sqrt{N} \simeq 32$, with maximum probability for $N=400$ of around 0.23. This can be increased as close to 1 as desired by standard
amplification techniques (repeating the search a few times). We see that subsequent peaks occur at other integer multiples of this initial time, $t=n(\pi/2)\sqrt{N}$ where $n=2,3,4.....$. 
%------------------%
%------------------%

As we have now shown how the quantum walk can be turned into a search algorithm, we are interested in how quickly the quantum walker finds the marked state. That is, we want to know when the probability of the walker being present at the marked state is at a maximum. As this probability is periodic and we want the algorithm to be efficient, the subsequent peaks are not of interest: we want to know when the first peak occurs. Although it would be ideal to measure the walker at the precise timing of the maximum in the first peak, this is not strictly necessary. In fact, as can be seen in fig.~\ref{maxprob2d}, the peaks are quite
broad, so even if an error occurs in when to measure, it only means a somewhat lower
probability of finding the marked state, this is only a constant
extra overhead on the amplification. For example, if the state of the walker was measured at half the optimal number of timesteps ($t=(\pi/4)\sqrt{N} \simeq 16$), the probability of the walker being measured in the marked state is roughly half that of the maximum possible ($p\approx0.1$). 

\begin{figure}[!tb]
    \begin{minipage}{\columnwidth}
	\centering
	\includegraphics[width=0.7\textwidth]{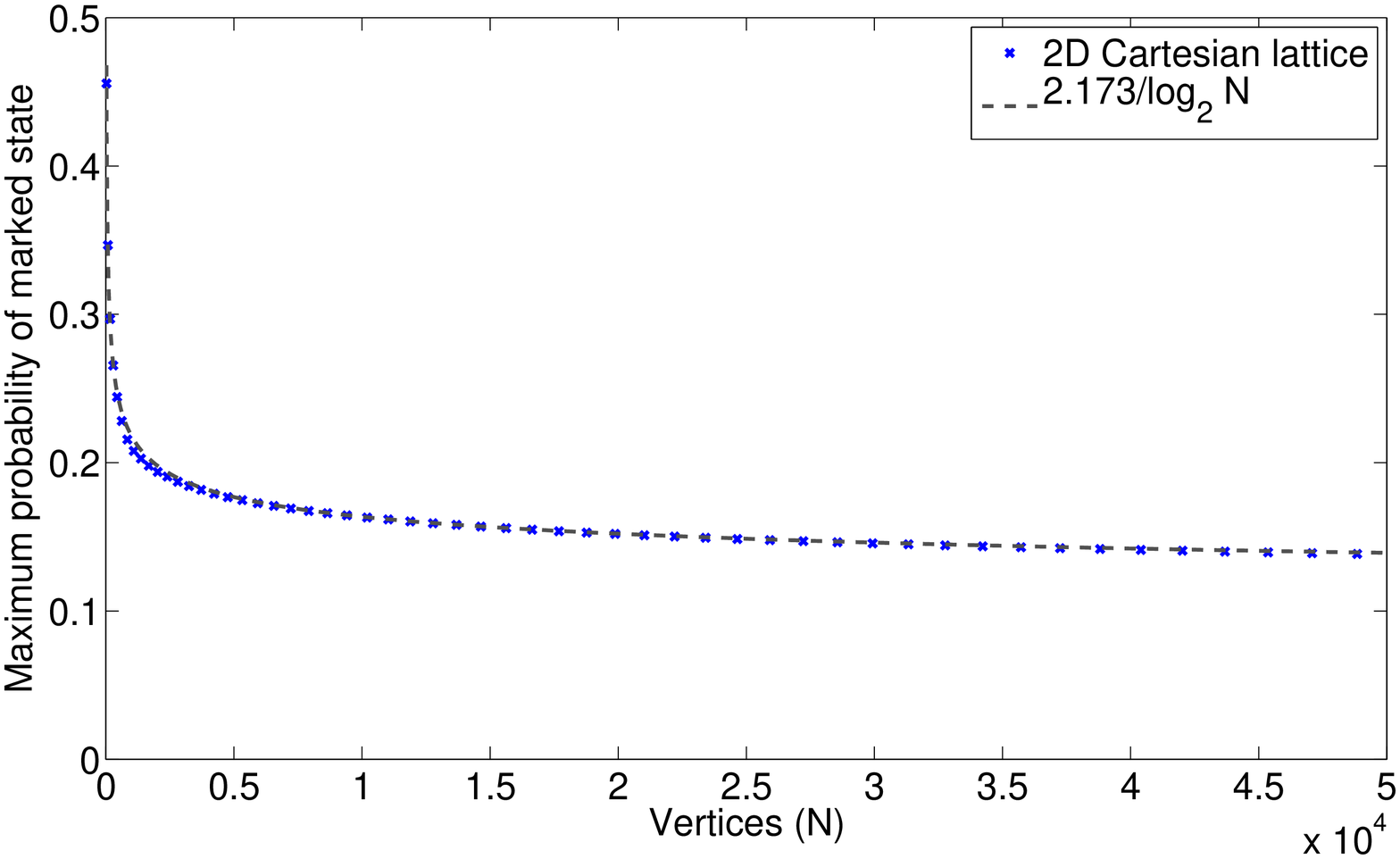}
	\caption[Plot to show how the maximum probability of the marked state varies with the size of the dataset on a 2D Cartesian lattice.]{Maximum of the first peak in the probability
	of being at the marked state for
	different sized data sets, using the optimal marked state coin
	in eq.~(\ref{markedcoin2d}) on a 2D lattice of size
	$\sqrt{N}\times\sqrt{N}$, plotted against $N$
	(crosses).  Also shown is the closest fit to our data, $2.173 / \log_{2} N$ (dashes).}
	\label{maxprob2dsizea}
    \end{minipage}
\end{figure}

\begin{figure}[!tb]
    \centering
	\includegraphics[width=0.7\textwidth]{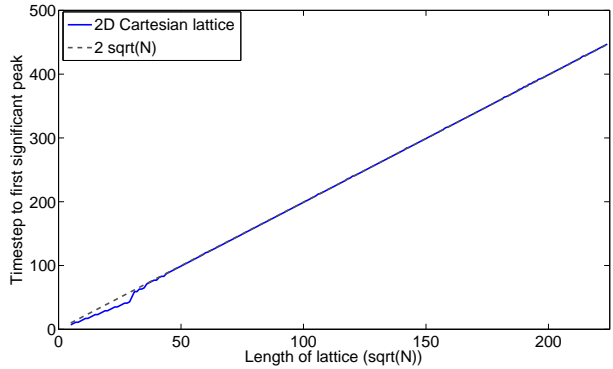}
	\caption[Plot to show how the time to find the maximum probability of the marked state varies with the size of the dataset on a 2D Cartesian lattice.]{Time step at which the first peak in the probability
	of being at the marked state occurs for different sized data sets,
	using the optimal marked state coin in eq.~(\ref{markedcoin2d}),
	plotted against $\sqrt{N}$. Also shown is the closest fit to our data, $2\sqrt{N}$.}
    \label{maxprob2dtime}
\end{figure}

In later sections, we discuss the algorithmic efficiency of the search algorithm on various graph structures. It is important we define here what factors of efficiency we are interested in. As we are looking to find a specific item from a set of many, we must consider how likely it is the walker coalesces at the marked state. The maximum probability of the walker at the marked state, i.e.~the maximum value of the first peak, varies with the size of the dataset (for the 2D Cartesian lattice). In this case, the theoretical value of $O(1/\log_2 N)$ from Ambainis \cite{ambainis04b} is numerically confirmed in our results in fig.~\ref{maxprob2dsizea} with a small prefactor of just over 2. The second factor we are interested in is the number of timesteps it takes to reach this maximum probability. The scaling of the time to find the marked state with the size of the dataset, $N$, for the 2D Cartesian lattice is shown in fig.~\ref{maxprob2dtime}. We see a scaling of $O(\sqrt{N})$ here, also with a prefactor of 2. 
%\hfill
%------------------%
%------------------%
%------------------%
%------------------%

In order to compare our results in later chapters to previous work, we must consider the total algorithmic complexity of the quantum walk search algorithm. In the case of the 2D Cartesian lattice, the maximum probability scales as $O(1/\log_2 N)$. Hence, we must use amplitude amplification techniques to increase this to a constant value. This has previously been shown to take $O(\sqrt{\log N})$ time steps \cite{brassard02a}. This makes the total algorithmic complexity $O(\sqrt{N \log N})$ for the 2D lattice, in agreement with the recent results of Tulsi \cite{tulsi08a} and Magniez et al. \cite{magniez08a}. These scalings are not the same for all graph structures. In particular, on a cubic lattice, the maximum probability scales as a constant value $O(1)$. As such, only a constant number of amplification steps are needed to bring the probability to $\approx 1$, thus the total algorithmic complexity is just $O(\sqrt{N})$.

%%%%%%%%%%%%%%%%%%%%%%%%%%%%%%%%%%%%
\section{Dimensionality}
\label{sec:dimensionality}
%%%%%%%%%%%%%%%%%%%%%%%%%%%%%%%%%%%%

In this section, we investigate how the spatial dimension of the database arrangement affects the searching algorithm. We already know, as noted in sec.~\ref{sec:searchalgorithm}, that the basic scaling of the algorithm is heavily dependent upon the spatial dimension of the structure in question. Both the scaling of the maximum probability of the marked state and the time to find this probability alter for structures of differing spatial dimension. We are interested here in how this basic scaling changes when the spatial dimension is altered. We investigate this in two ways: firstly by introducing a simple form of tunnelling, which allows us to interpolate between structures of varying spatial dimension, and secondly by using lattices of varying height (1D-2D) and depth (2D-3D). 

%%%%%%%%%%%%%%%%%%%%%%%%%%%%%%%%%%
\subsection{Tunnelling operator}
%%%%%%%%%%%%%%%%%%%%%%%%%%%%%%%%%%

\begin{figure}[!bt]
\begin{minipage}{\columnwidth}
\centering
\includegraphics[width=0.25\textwidth]{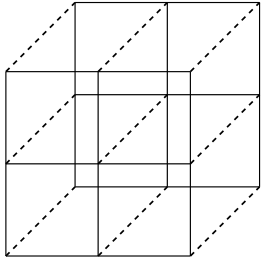}
\caption[Structure used to interpolate between a 2D and a 3D lattice.]{Basic unit of the structure we use to interpolate between the two dimensional Cartesian and the three dimensional cubic lattices. The solid lines represent the fixed, normal edges whereas the dashed lines represent the edges we set to be tunnelling.}
\label{2d3dfig}
\end{minipage}
\end{figure}

We now describe a modified coin operator we will use in the search algorithm to gradually vary the connectivity of the structures studied. We introduce a simple form of tunnelling to allow us to gradually vary the substrate to be walked upon from one form to another. A simple example is changing a set of 2D Cartesian lattices into a cubic lattice by introducing connecting links between the lattices, see fig.~\ref{2d3dfig}.
In order to achieve the quantum walk dynamics we require, we must use a different coin operator. The only condition on this operator is that it must be unitary. As such, we `design' a new coin operator which incorporates a single tunnelling parameter, $c$, which will allow us to vary the strength of specific tunnelling edges. We define $d$ to be the degree of the vertex in question as used previously in the Grover coin, eq.~(\ref{grovercoin}), and $t$ to be the number of tunnelling edges. For a $d$-dimensional vertex, the first ($d-t$) edges in our labelling scheme are normal and the last $t$ edges are tunnelling. In fig.~\ref{2d3dfig}, the solid edges are normal, fixed edges creating a 2D lattice and the dashed edges are tunnelling edges which convert the 2D lattices to a cubic one. The general matrix for the desired coin operator would be as follows:
\begin{equation}
T_{d,t} = \left( \begin{array}{cccccccccc} a & b & b & \dots & b & c & c & c & \dots & c\\
b & a & b & \dots & b & c & c & c & \dots & c\\
. & . & . & \dots & . & . & . & . & \dots & .\\
. & . & . & \dots & . & . & . & . & \dots & .\\
. & . & . & \dots & . & . & . & . & \dots & .\\
b & b & b & \dots & a & c & c & c & \dots & c\\
c & c & c & \dots & c & e & f & f & \dots & f\\
c & c & c & \dots & c & f & e & f & \dots & f\\
. & . & . & \dots & . & . & .  & . & \dots & .\\
. & . & . & \dots & . & . & . & . & \dots & .\\
. & . & . & \dots & . & . & . & . & \dots & .\\
c & c & c & \dots & c & f & f & f & \dots & e
 \end{array} \right )
\label{tunnellingmatrix}
\end{equation}
where the blocks of ``$abbb...$'' are $d-t$ square and the blocks of ``$efff...$'' are $t$ square. We want to be able to rewrite this in terms of just one tunnelling parameter, $c$, which represents the coupling between the normal and tunnelling edges. As the dynamics of the walk must be reversible, we must ensure that the coin produced is unitary. As such, 
\begin{equation}
\left ( T_{d,t} \right ) \left ( T_{d,t} \right )^{\dagger} = \mathbb{I}_{d},
\label{unitarity}
\end{equation}
must hold, where $\left ( T_{d,t} \right )^{\dagger}$ is the hermitian conjugate of the general matrix. We can solve the five equations which are formed as a consequence in terms of just the degree of the lattice, $d$, the number of tunnelling edges, $t$ and the tunnelling parameter, $c$, giving
\begin{equation}
a=b-1,
\label{solveda}
\end{equation}
\begin{equation}
b = \frac{1+ \sqrt{1 - (d-t)tc^2}}{d-t},
\label{solvedb}
\end{equation}
\begin{equation}
e=f-1,
\label{solvede}
\end{equation}
and
\begin{equation}
f = \frac{1-\sqrt{1 - (d-t)tc^2}}{t}.
\label{solvedf}
\end{equation}

This operator allows us to vary the `strength' of certain edges in the structure we wish to walk on. It holds for any degree of vertex but the number of tunnelling edges must be at most half the degree, i.e. $t\leq d/2$. By using vertices of degree six with two tunnelling edges, we have a set of basic 2D Cartesian lattices gradually becoming a cubic lattice as in fig.~\ref{2d3dfig}. In this case, setting $c=0$ we obtain
\begin{equation}
T_{6,2} = \frac{1}{2} \left ( \begin{array}{cccccc} -1 & 1 & 1 & 1 & 0 & 0\\ 1 & -1 & 1 & 1 & 0 & 0 \\ 1 & 1 & -1 & 1 & 0 & 0\\ 1 & 1 & 1 & -1 & 0 & 0\\ 0 & 0 & 0 & 0 & -2 & 0\\ 0 & 0 & 0 & 0 & 0 & -2 \end{array} \right ),
\label{tunneling62c0}
\end{equation}  
and setting $c=2/d=1/3$ gives
\begin{equation}
T_{6,2} = \frac{1}{3} \left ( \begin{array}{cccccc} -2 & 1 & 1 & 1 & 1 & 1\\ 1 & -2 & 1 & 1 & 1 & 1\\ 1 & 1 & -2 & 1 & 1 & 1\\ 1 & 1 & 1 & -2 & 1 & 1\\ 1 & 1 & 1 & 1 & -2 & 1\\ 1 & 1 & 1 & 1 & 1 & -2 \end{array} \right ).
\label{tunneling62cd}
\end{equation}
These choices of $c$ represent the extremes of the operator when the structure would be either a basic 2D Cartesian lattice, eq.~(\ref{tunneling62c0}), or a cubic lattice, eq.~(\ref{tunneling62cd}). Any other values of $c$ where $0\le c\le 2/d$ would give varying strengths of tunnelling across the tunnelling edges.

\subsection{Interpolating between the 2D and 3D lattices using the tunnelling operator}

Using the tunnelling operator introduced above, we numerically study how the algorithm is affected by the change in spatial dimension. We use the operator to interpolate between a Cartesian lattice (2D) and a cubic lattice (3D). In this case, we interpolate between a set of 2D Cartesian lattices and a fully connected cubic lattices by using vertices of degree six with two tunnelling edges, with the correct connectivity, fig.~\ref{2d3dfig}. 

Although we do not show the results here, we also performed the same investigation between the 1D line and the 2D lattice, finding a quantitatively similar result. The results in this case were more difficult to analyse due to the search algorithm failing as the structure approached a line.

The search algorithm we use is the same as in the original work by Shenvi, Kempe and Whaley \cite{shenvi02a}, with a small change to the initial state. Although we must still start the walker in a uniform superposition over all vertices, the distribution over the edges must be altered slightly to account for the strength of the tunnelling edges. In other words, we must distribute the state over the edges with a weighting to match the tunnelling strength as follows
\begin{equation}
(d-t) \alpha + tp \alpha = \frac{1}{\sqrt{N}},
\label{initialspreadtunnelling}
\end{equation}
where $p$ is the tunnelling probability, $\alpha$ is the state on each edge and $N$ is the number of vertices. The tunnelling probability is just the tunnelling parameter, $c$, rescaled to lie between $0$ and $2/d$. In this way, the tunnelling probability matches the proportion of the initial state which is placed on the tunnelling edges. This initial state gives a probability distribution, where there is no marked state, which is periodic over two timesteps as can easily be checked. Although this is not stationary as in the case of the basic non-tunnelling lattices, the fact that it returns to the same state after only two timesteps means it will give rise to the same dynamics.

We ran the algorithm on a 3D cubic lattice where the edges which link the `slices' of 2D lattices together are tunnelling. We show the a basic unit in fig.~\ref{2d3dfig}. The solid lines are the fixed, normal edges and the dashed lines are tunnelling edges. This structure allows us to gradually change the strength of the edges which make the structure three dimensional, hence interpolating between the 2D Cartesian lattice and the 3D cubic lattice.

\begin{figure}[!tb]
\begin{minipage}{\columnwidth}
\centering
\includegraphics[width=0.75\textwidth]{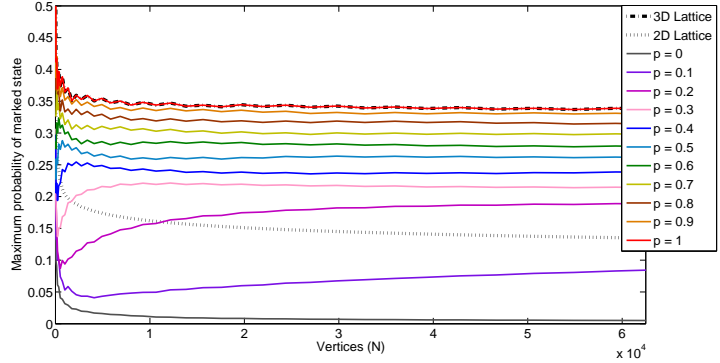}
\caption[Plot to show how the maximum probability of the marked state varies with the size of the dataset and the tunnelling strength (2D-3D).]{Plot to show how the maximum probability of the marked state varies with both the size of the lattice and the tunnelling strength as a two dimensional lattice is gradually changed into a three dimensional Cartesian lattice.}
\label{2d3dprob}
\end{minipage}
\end{figure}

The maximum probability of the marked state, shown in fig.~\ref{2d3dprob}, changes from the $1/\log_{2} N$ scaling in the 2D case to the constant $O(1)$ scaling as soon as the additional edges even have a small weighting attached to them. 
\begin{figure}[!tb]
\begin{minipage}{\columnwidth}
\centering
\includegraphics[width=0.75\textwidth]{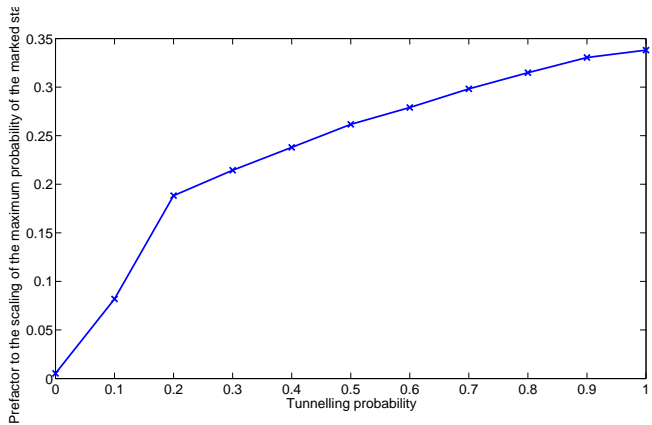}
\caption[Plot to show how the prefactor to the scaling of the maximum probability of the marked state varies with tunnelling strength (2D-3D).]{Plot to show how the prefactor to the scaling of the maximum probability of the marked state, obtained from the data in fig.~\ref{2d3dprob}, varies with the tunnelling strength as a two dimensional Cartesian lattice is gradually changed into a three dimensional lattice.}
\label{2d3dprobscaling}
\end{minipage}
\end{figure}
At low tunnelling strengths, we see the probability dropping initially before gradually recovering towards a constant value for higher lattice sizes. At these higher sizes, it is easy to see that the scaling is constant for any tunnelling strength with just varying prefactors. Figure \ref{2d3dprobscaling} shows how this prefactor to the scaling of the maximum probability of the marked state changes as we increase the tunnelling probability. The sharp drop at the low tunnelling probabilities is most probably due to the fact the scaling hasn't reached the constant value as we can only simulate up to a fixed lattice size.

In addition, we note here that when the probability of tunnelling is zero, the scaling does not match that of the basic 2D lattice. At $p=0$, the structure is in effect a collection of 2D lattices which are unlinked. The initial state is still spread across all these individual lattices and due to the connectivity of the structure, only the amplitude in one of the lattices (that with the marked state present) is able to coalesce on the marked state. As such, the scaling is just reduced by a constant factor as can be seen in fig.~\ref{2d3dprob}. 

\begin{figure}[!tt]
\begin{minipage}{\columnwidth}
\centering
\includegraphics[width=0.75\textwidth]{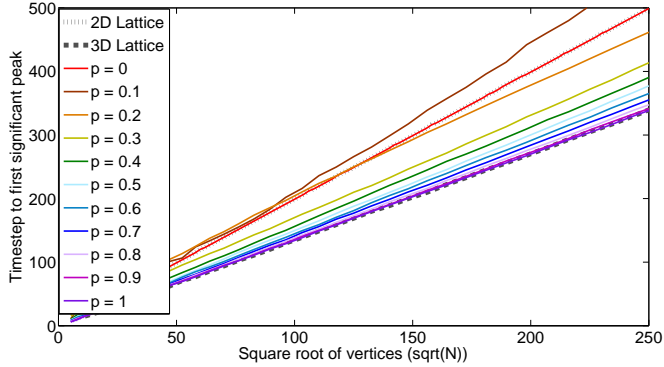}
\caption[Plot to show how the time to find the maximum probability of the marked state varies with the size of the dataset and the tunnelling strength (2D-3D).]{Plot to show how the time to find the marked state varies with both the size of the lattice and the tunnelling strength as a two dimensional lattice is gradually changed into a three dimensional Cartesian lattice.}
\label{2d3dtime}
\end{minipage}
\end{figure}

The time to find the marked state follows a similar behaviour. We firstly show, fig.~\ref{2d3dtime}, how the scaling of the time to find the marked state varies with both the size of the lattice and the tunnelling strength.
\begin{figure}[!tt]
\begin{minipage}{\columnwidth}
\centering
\includegraphics[width=0.75\textwidth]{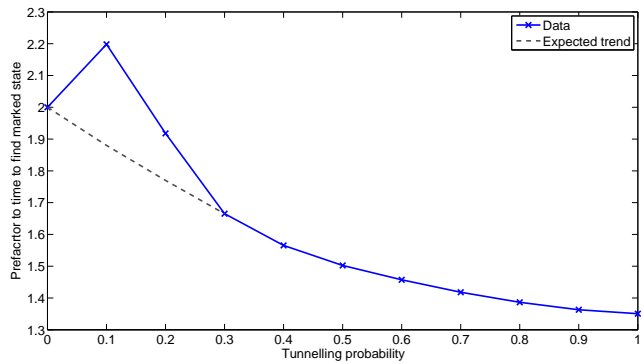}
\caption[Plot to show how the prefactor to the scaling of the time to find the maximum probability of the marked state varies with tunnelling strength (2D-3D).]{Plot to show how the prefactor to the scaling of the time to find the marked state, obtained from the data in fig.~\ref{2d3dtime}, varies with the tunnelling strength as a two dimensional lattice is gradually changed into a three dimensional Cartesian lattice.}
\label{2d3dtimescaling}
\end{minipage}
\end{figure}
The basic scaling of the time to find the marked state is the same in both two and three dimensions, $O(\sqrt{N})$. We see that, in general, the time to find the marked state (the prefactor to the basic scaling) decreases as the tunnelling strength increases, thus making the algorithm more efficient. We show a plot of how this prefactor varies with the tunnelling strength in fig.~\ref{2d3dtimescaling}. We do note that at the very low tunnelling probabilities, $p<\approx 0.3$, the scaling of the time to find the marked state does not follow the same behaviour. This is due to the fact that the probability at the marked state has not yet reached a constant value, as mentioned previously. If we were able to run the algorithm on much larger sized lattices, we should find the probability of the marked state stabilising and thus the time to find the marked state matching the quadratic speedup in scaling. We show the expected trend to the scaling of the prefactors to the time to find the marked state in fig.~\ref{2d3dtimescaling}.

\subsection{Interpolating between the 2D and 3D lattices by varying the depth}

In this section, we discuss lattices of varying depth to give a different avenue of investigation of the dependence on spatial dimension. At low depths, a 3D lattice can be viewed in effect as a 2D lattice. We are interested in how the scaling of the probability of the marked state and the time to find it changes as the lattice depth is gradually increased, eventually becoming a fully symmetric 3D lattice. As with the results of the tunnelling operator, we also investigated this interpolation between the 1D line and the 2D lattice showing a quantitatively similar behaviour. Again, due to the algorithm failing as the structure approaches the line, it is much clearer to see the behaviour in the 2D-3D case.
\begin{figure}[!tb]
\begin{minipage}{\columnwidth}
\centering
\includegraphics[width=0.75\textwidth]{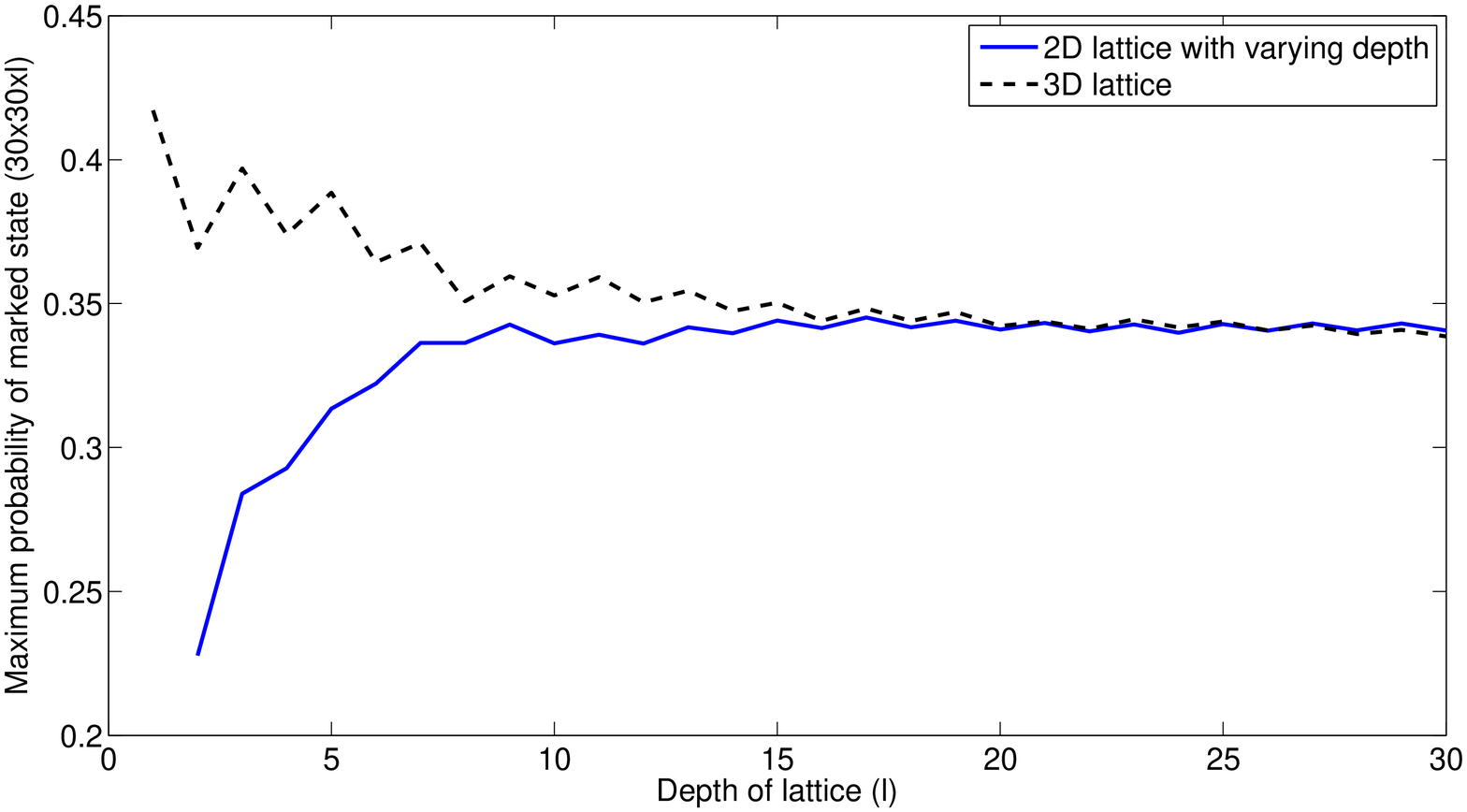}
\caption[Plot to show how the maximum probability of the marked state varies with the depth of the lattice for a fixed width and height.]{Maximum probability of the marked state as the depth of the 3D lattice is increased from one layer to a perfect cube of 30x30x30 sites (solid line). Maximum probability of the marked state for cubic lattices of varying size is shown for comparison (dashed line).}
\label{fixedprob2d3d}
\end{minipage}
\end{figure}

We firstly show how the maximum probability of the marked state varies for a fixed width and height of lattice (30x30), while varying the depth. Figure \ref{fixedprob2d3d} shows how this probability varies, along with the equivalent scaling for the fully cubic 3D lattice, showing the scaling matches at roughly $l=15$, where $l$ is the depth of the lattice.

\begin{figure}[!bt]
\begin{minipage}{\columnwidth}
\centering
\includegraphics[width=0.75\textwidth]{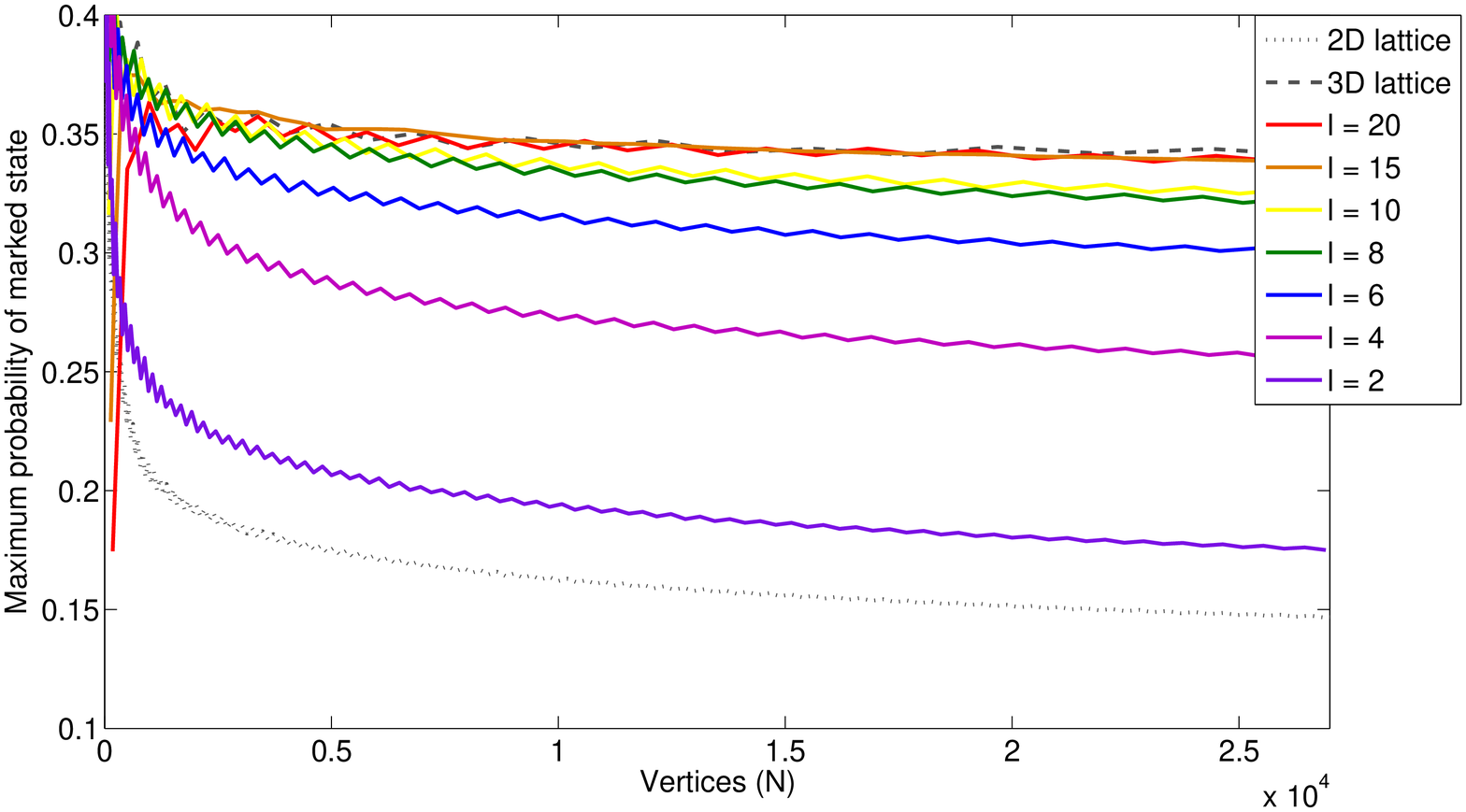}
\caption[Plot to show how the maximum probability of the marked state varies with the depth of the lattice.]{Plot to show how the maximum probability of the marked state varies as a two dimensional lattice is gradually increased in depth (l) to become a three dimensional cubic lattice. We maintain the same number of vertices in each case.}
\label{prob2d3dmxn}
\end{minipage}
\end{figure}

We then fixed the depth of the lattice and altered the width and height of the lattice (with a fixed number of vertices) for each run of the search algorithm. Figure \ref{prob2d3dmxn} shows how the maximum probability of the marked state varies for differing depths of the lattice. We again see a gradual change in scaling from the basic 2D logarithmic scaling to the constant scaling of the cubic lattice. This is in contrast to the almost instantaneous change in scaling we see when interpolating using the tunnelling operator.

The basic scaling of the time to find the marked state is unaffected by the change in the depth of the lattice. The prefactor to this scaling though does decrease as we increase the depth, changing from that of the basic 2D lattice to almost match that of the cubic lattice for even lattices of modest depth. \begin{figure}[!bt]
\begin{minipage}{\columnwidth}
\centering
\includegraphics[width=0.75\textwidth]{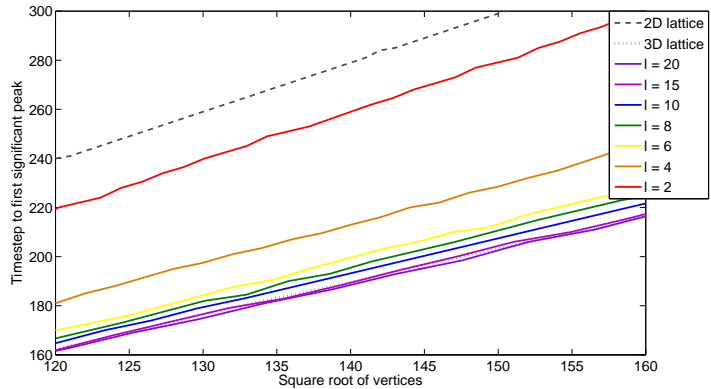}
\caption[Plot to show how the time to find the maximum probability of the marked state varies with the depth of the lattice.]{Plot to show how the time to find the marked state varies as a two dimensional lattice is gradually increased in depth (l) to become a three dimensional cubic lattice. We maintain the same number of vertices in each case.}
\label{time2d3dmxn}
\end{minipage}
\end{figure}

%%%%%%%%%%%%%%%%%%%%%%%%%%%%%%%%%%%%%%%%%%
\section{Connectivity}
\label{sec:connectivity}
%%%%%%%%%%%%%%%%%%%%%%%%%%%%%%%%%%%%%%%%%%

We now investigate how important the connectivity of the database arrangement is for the searching algorithm. As previously discussed, the basic scaling of the algorithm is heavily dependent upon the spatial dimension of the structure in question. Here, we numerically study how the connectivity, in a specific spatial dimension, affects the prefactors to this scaling. Although it is unlikely that the runtime of the search algorithm on a 2D Cartesian lattice can be reduced to the optimal $O(\sqrt{N})$, it may be possible to reduce any constant overhead associated with the run time. 

We use the tunnelling coin operator introduced in the previous section which allows us to model the search algorithm on structures where there is a probability of additional connections existing. For example, our tunnelling operator allows us to interpolate between running the search algorithm on a hexagonal lattice, with degree $d=3$, and the 2D Cartesian lattice, $d=4$. This allows us to analyse how the search algorithm is affected by a gradual change in the degree of the underlying substrate by changing the tunnelling strength of the additional edges. This extends the initial studies of \cite{lovett10b} by considering an interpolation between lattices with fixed degree. We firstly introduce the structures we wish to perform the search algorithm upon, before presenting our results for both two and three dimensional structures with varying connectivity. We show that the prefactors to the scaling of the algorithm for both the maximum probability of the marked state and the time to find the marked state are dependent on the connectivity of the underlying structure. 

\subsection{Two dimensional structures}

\begin{figure}[!bt]
\begin{minipage}{\columnwidth}
\centering
\includegraphics[width=0.8\textwidth]{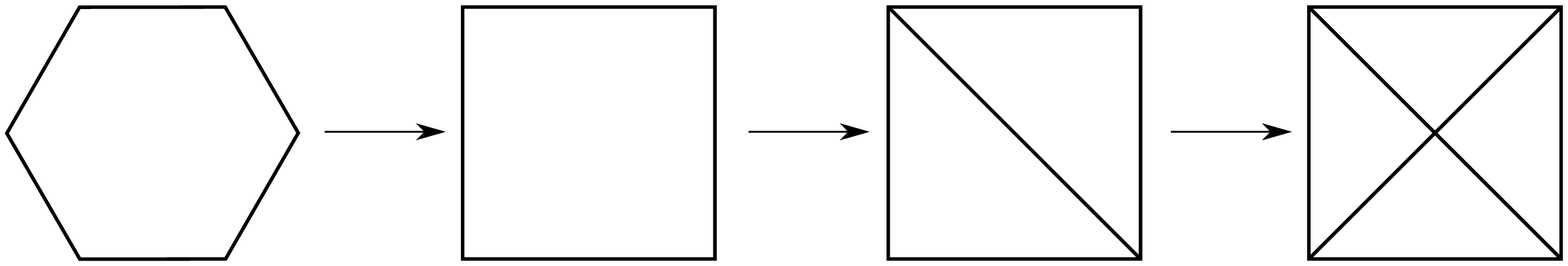}
\caption[The 2D lattices we interpolate between using the tunnelling operator.]{The 2D lattices we interpolate between using the tunnelling matrix. We change gradually from a hexagonal lattice, $d=3$, to a 2D Cartesian lattice with diagonals included, $d=8$. We show here just the building block of each lattice. We note that in the case of the 2D Cartesian lattice with diagonals, there is no vertex at the central point where the edges cross.}
\label{2dlattices}
\end{minipage}
\end{figure}

Using the tunnelling matrix we have introduced, we ran the search algorithm on 2D lattices ranging from $d=3$, a hexagonal lattice, through to $d=8$, a Cartesian lattice with diagonals added as shown in fig.~\ref{2dlattices},
for varying lattices sizes from $6^{2}$ (36) vertices up to $250^{2}$ (62500) vertices. As in fig.~\ref{2dlattices}, we gradually changed the degree of the structure we performed the search algorithm on. This was split into intermediate steps, firstly from the 2D hexagonal lattice ($d=3$) to the square lattice ($d= 4$), the square lattice to the triangular lattice ($d=6$), eventually ending at the more highly connected Cartesian lattice with diagonals ($d=8$). We spread the walker in the same fashion as eq.~(\ref{initialspreadtunnelling}) to ensure we distribute the state evenly based on the tunnelling strength of the edges.

\begin{figure}[!bt]
\begin{minipage}{\columnwidth}
\centering
\includegraphics[width=0.75\textwidth]{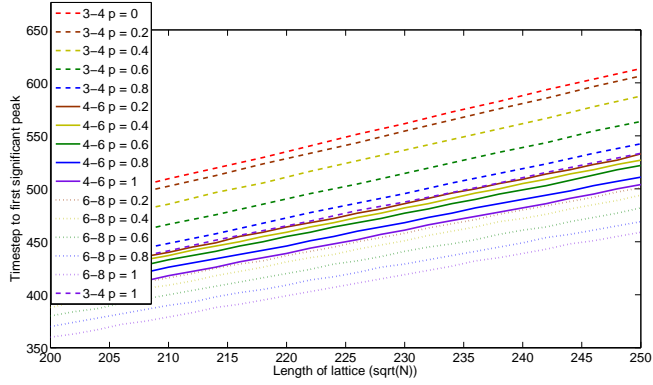}
\caption[Scaling of the time to find the marked state with the size of the lattice with varying connectivity in two dimensions.]{Plot to show how the time to find the marked state varies with both the size of the lattice and varying connectivity in two dimensions. It is clear that as the connectivity of the structure increases, the time to find the marked state decreases. Note that this is a zoomed in plot showing only larger lattices sizes, data for $\sqrt{N} < 200$  has been omitted to improve clarity.}
\label{2dtimeall}
\end{minipage}
\end{figure}

\begin{figure}[!bt]
\begin{minipage}{\columnwidth}
\centering
\includegraphics[width=0.75\textwidth]{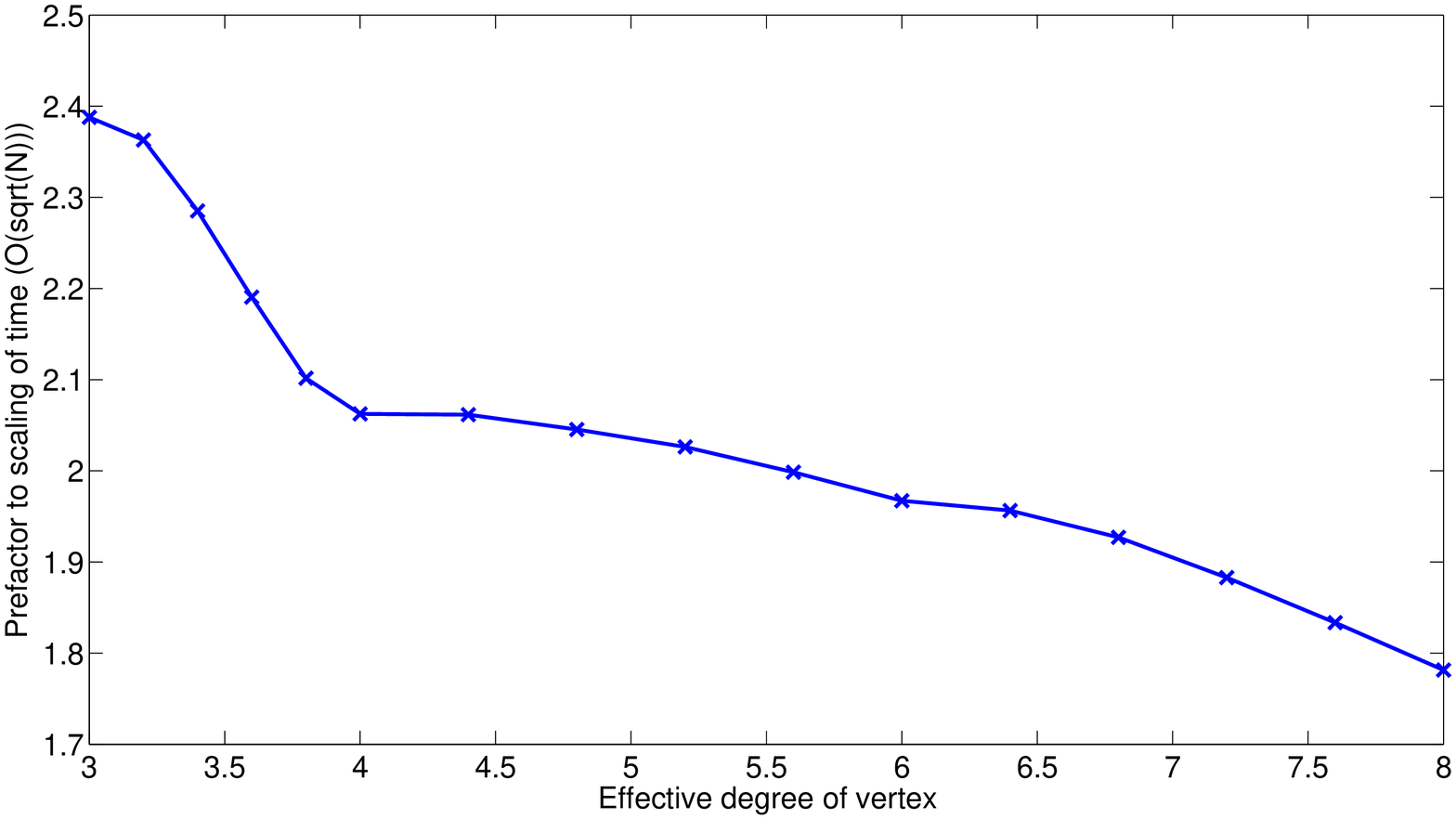}
\caption[Plot to show how the prefactor to the scaling of the time to find the marked state varies with the connectivity in two dimensions.]{Plot to show how the prefactor to the scaling, obtained from the data shown in fig.~\ref{2dtimeall}, of the time to find the marked state of $O(\sqrt{N})$ changes with the degree of the two dimensional structure being searched.}
\label{2dtimeprefactor}
\end{minipage}
\end{figure}

We show in fig.~\ref{2dtimeall} how the time to find the marked state varies with both the size of the lattice and the connectivity.
We see that as the connectivity increases, the time to find the marked state decreases, hence the efficiency of the algorithm increases. As the time to find the marked state scales as $O(\sqrt{N})$, we fit to each of the data sets in fig.~\ref{2dtimeall} to obtain the prefactor to the scaling of the time to find the marked state. Figure \ref{2dtimeprefactor} shows how this prefactor to the scaling changes with the degree of the underlying structure being searched. We see although there is no specific scaling here, we do note there seems to be a symmetry effect for integer degree, though less strong than might have been expected.

\begin{figure}[!bt]
\begin{minipage}{\columnwidth}
\centering
\includegraphics[width=0.75\textwidth]{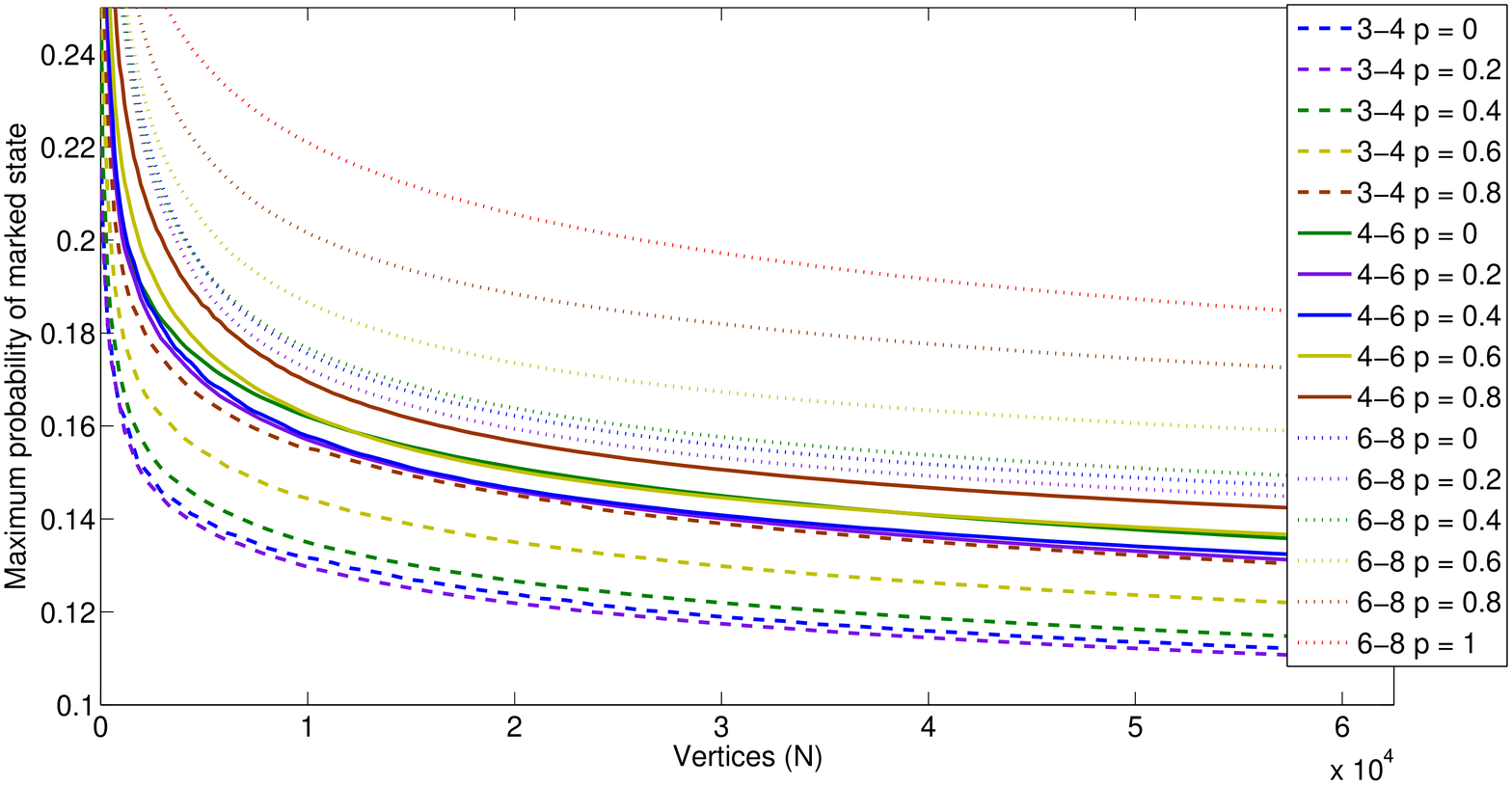}
\caption[Scaling of the maximum probability of the marked state with the size of the lattice with varying connectivity in two dimensions.]{Plot to show how the maximum probability of the marked state varies with both the size of the lattice and varying connectivity in two dimensions. In general, as the connectivity of the structure increases, the maximum probability of the marked state also increases.}
\label{2dproball}
\end{minipage}
\end{figure}

\begin{figure}[!bt]
\begin{minipage}{\columnwidth}
\centering
\includegraphics[width=0.75\textwidth]{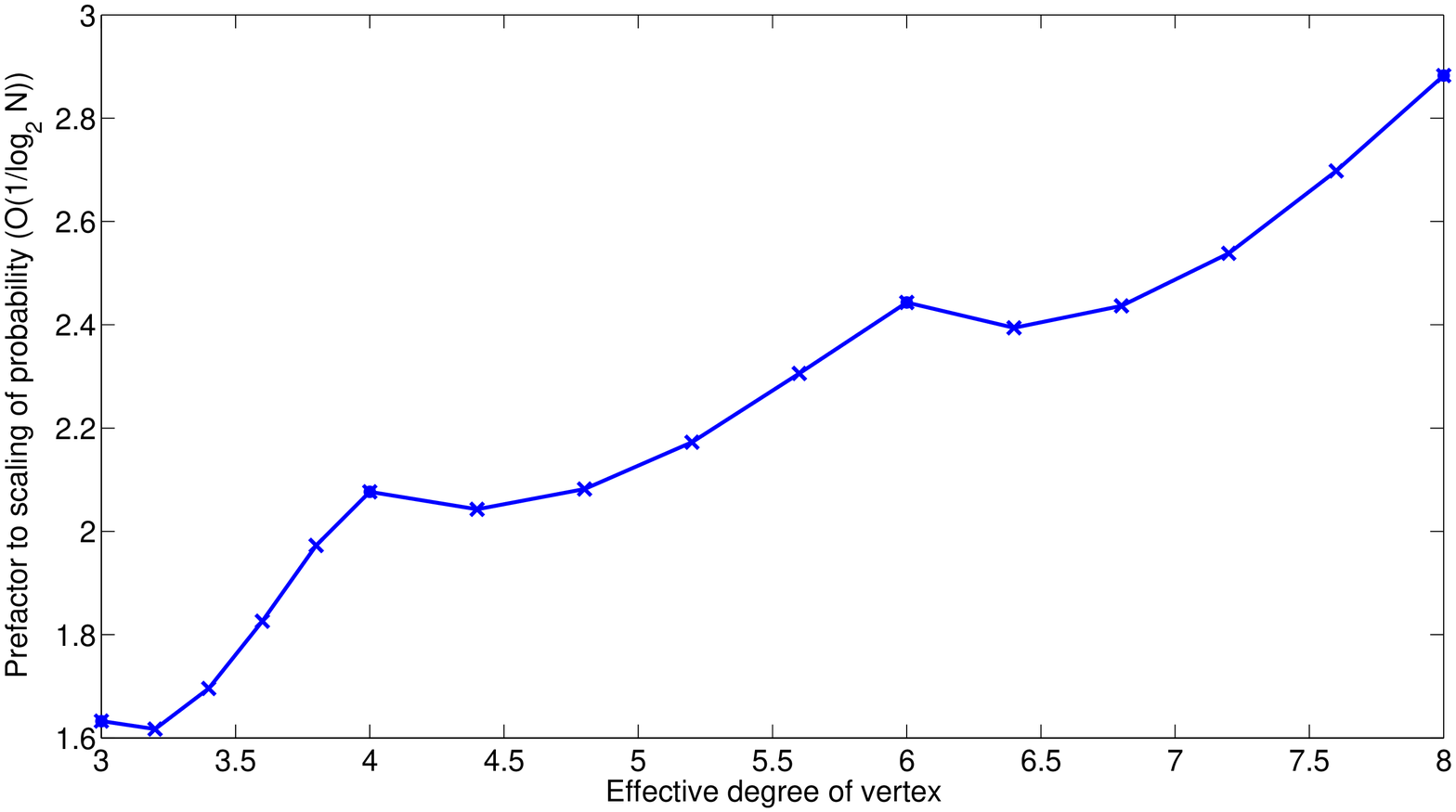}
\caption[Plot to show how the prefactor to the scaling of the maximum probability of the marked state varies with the connectivity in two dimensions.]{Plot to show how the prefactor to the scaling, obtained from the data in fig.~\ref{2dproball}, of the maximum probability of the marked state of $O(1/\log_{2} N)$ changes with the degree of the two dimensional structure being searched.}
\label{2dprobprefactor}
\end{minipage}
\end{figure}

In fig.~\ref{2dproball}, we show how the maximum probability of the marked state varies with both the size of the dataset and the connectivity of the structure. We see that, in general, as the connectivity of the structure being searched increases, the maximum probability of the marked state also increases. 
A larger prefactor to this scaling means fewer repeats of the algorithm are required to bring the success probability close to unity. Figure \ref{2dprobprefactor} shows how this prefactor to the scaling of $O(1/\log_{2} N)$ varies with the degree of the structure being searched.
The `dips' and revivals in the scaling seem counter intuitive, but appear to arise from the dynamics of the walk on these structures where the symmetry is partially broken (low tunnelling strength). In order to confirm this, we briefly examined the basic dynamics of the quantum walk while varying the tunnelling strength. We started the walker at a specific vertex in the graph, as opposed to an equal superposition, and allowed it to propagate outwards in order to determine its dynamics. We define the spread of the walker as
\begin{equation}
\langle r\rangle = \sum^{N}_{i=1} p_{i} s_{i},
\label{spreadeq}
\end{equation}
where $p_{i}$ is the probability of the walker being at vertex $i$ and $s_{i}$ is the shortest path distance from the position of the initial state to vertex $i$. Using this metric for the rate of spreading, we explored how this was affected by the tunnelling strength. 
\begin{figure}[!bt]
\begin{minipage}{\columnwidth}
\centering
\includegraphics[width=0.75\textwidth]{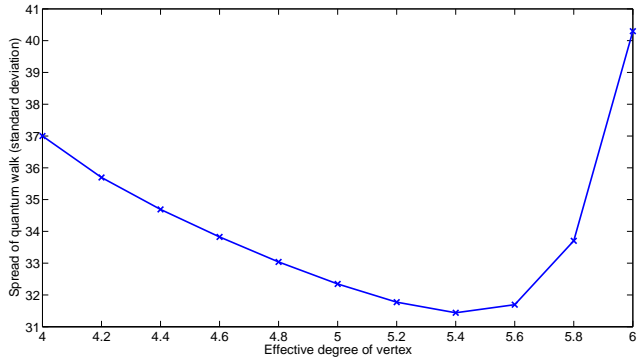}
\caption[Plot to show how the spread of the quantum walk varies with the connectivity in two dimensions.]{Plot to show how the spreading of the quantum walk, characterised by eq.~(\ref{spreadeq}), changes with the degree of the two dimensional structure being searched. In this case, we show the dynamics of the spread of the quantum walk on a two dimensional Cartesian lattice ($d=4$) being changed into a triangular lattice ($d=6$)}
\label{spread}
\end{minipage}
\end{figure}
Figure \ref{spread} shows this spreading on a 2D Cartesian lattice gradually being turned into a triangular lattice as in fig.~\ref{2dlattices}. We found at low tunnelling strengths, where the symmetry breaking is most obvious, the spread, $\langle r\rangle$, dropped. As the tunnelling strength was raised, the quantum walk was able to recover and $\langle r\rangle$ increased back to the value of the original lattice, before increasing further as the tunnelling strength reached its maximum value, i.e. the new lattice. Although this is not an exhaustive study of the quantum walk dynamics when we include tunnelling edges, this behaviour does match the results we find for the search algorithm. While the variation of $\langle r\rangle$ does not match the scaling of the probability of the marked state directly, the basic quantum walk dynamics do not have any reflection effects from the edges of the structure. Due to the periodic boundary conditions imposed in the searching algorithm, we find slightly different behaviour which relate to the extra interference effects. We found similar results for all the lattices studied in both two and three dimensions.

\begin{figure}[!hbt]
\begin{minipage}{\columnwidth}
\centering
\includegraphics[width=0.8\textwidth]{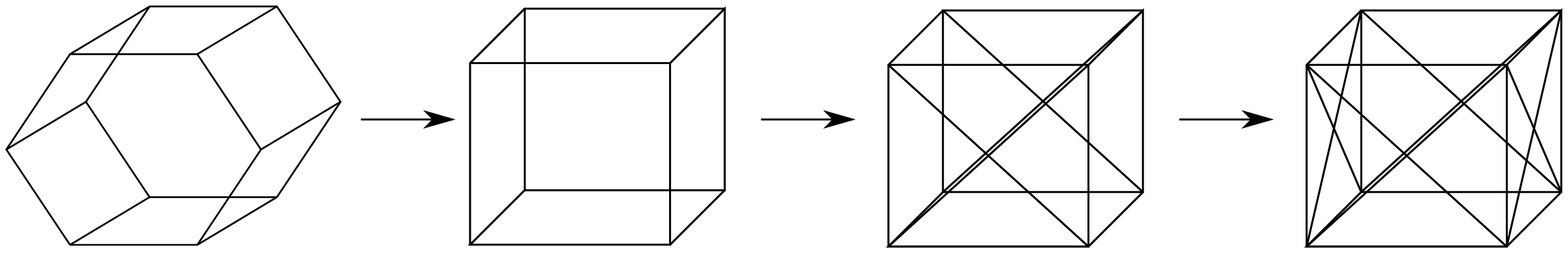}
\caption[The 3D lattices we interpolate between using the tunnelling operator.]{The 3D lattices we interpolate between using the tunnelling matrix. We change gradually from a 3D hexagonal lattice, $d=5$, through to a cubic lattice with diagonals added on the faces, $d=14$. We show here just the building block of each lattice. We note here that vertices are only present at the eight corners of the cubic structures, there are no vertices present where the edges cross.}
\label{3dlattices}
\end{minipage}
\end{figure}

\subsection{Three dimensional structures}

We now consider three dimensional lattices, using the tunnelling matrix to study structures ranging from $d=5$, a 3D hexagonal lattice, through to $d=14$, a cubic lattice with additional diagonals added as shown in fig.~\ref{3dlattices}.

We ran the search algorithm for varying lattices sizes from $3^{3}$ (27) vertices up to $40^{3}$ (64000) vertices. As in the two dimensional case, we did not just change the lattice from $d=5$ to $d=14$ in one go. We split this into intermediary steps, fig.~\ref{3dlattices}, changing firstly from the 3D hexagonal lattice ($d=5$) to the cubic lattice ($d= 6$), the cubic lattice to one with diagonals added to one face ($d=10$), eventually ending with a cubic lattice with diagonals added on two faces ($d=14$). We split the initial state across the vertices and edges in the same way as in the two dimensional case. 

\begin{figure}[!bt]
\begin{minipage}{\columnwidth}
\centering
\includegraphics[width=0.75\textwidth]{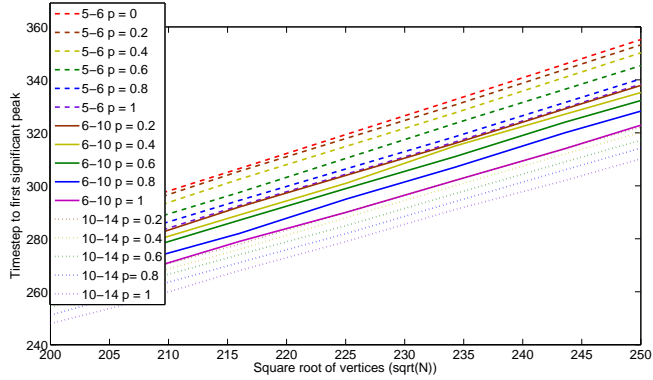}
\caption[Scaling of the time to find the marked state with the size of the lattice with varying connectivity in three dimensions.]{Plot to show how the time to find the marked state varies with the size of the lattice and varying connectivity in three dimensions. It is clear that as the connectivity of the structure increases, the time to find the marked state decreases. Note that this is a zoomed in plot showing large lattices sizes, data for $\sqrt{N} < 200$  has been omitted to improve clarity.}
\label{3dtimeall}
\end{minipage}
\end{figure}

We show in fig.~\ref{3dtimeall} how the time to find the marked state varies with both the size of the lattice and the connectivity.
\begin{figure}[!bt]
\begin{minipage}{\columnwidth}
\centering
\includegraphics[width=0.75\textwidth]{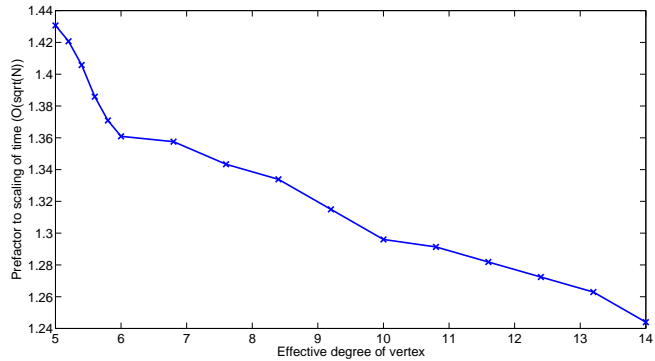}
\caption[Plot to show how the prefactor to the scaling of the time to find the marked state varies with the connectivity in three dimensions.]{Plot to show how the prefactor to the scaling, obtained from the data shown in fig.~\ref{3dtimeall}, of the time to find the marked state of $O(\sqrt{N})$ changes with the degree of the three dimensional structure being searched.}
\label{3dtimeprefactor}
\end{minipage}
\end{figure}
It is clear that as the connectivity increases, the time to find the marked state decreases, hence the efficiency of the algorithm increases. As the time to find the marked state scales as $O(\sqrt{N})$, we fit to each of these to obtain the prefactor to the scaling of the time to find the marked state. Figure \ref{3dtimeprefactor} shows how this prefactor to the scaling changes with the degree of the underlying structure being searched. 

\begin{figure}[!bt]
\begin{minipage}{\columnwidth}
\centering
\includegraphics[width=0.75\textwidth]{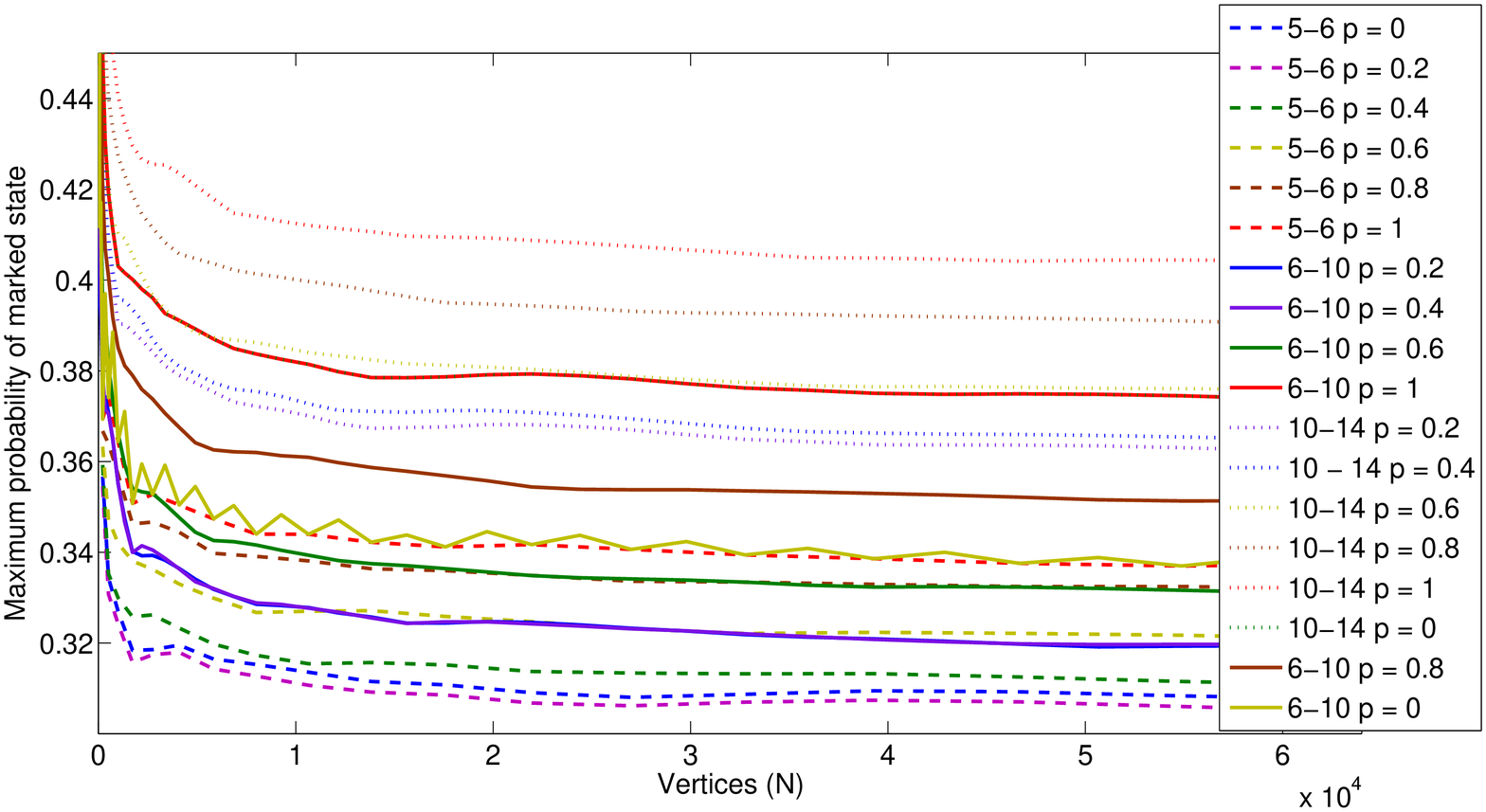}
\caption[Scaling of the maximum probability of the marked state with the size of the lattice with varying connectivity in three dimensions.]{Plot to show how the maximum probability of the marked state varies with the size of the lattice and varying connectivity in three dimensions. In general, as the connectivity of the structure increases, the maximum probability of the marked state also increases.}
\label{3dproball}
\end{minipage}
\end{figure}

In the three dimensional case, the maximum probability of the marked state scales in a constant fashion, $O(1)$. As such, this scaling does not affect the complexity of the algorithm as in the two dimensional case. However, we do note that this constant value of probability does affect how many times we must run the algorithm to ensure we have the correct result. We show in fig.~\ref{3dproball} that, in general, as the connectivity of the structure being searched increases, the maximum probability of the marked state also increases.
\begin{figure}[!bt]
\begin{minipage}{\columnwidth}
\centering
\includegraphics[width=0.75\textwidth]{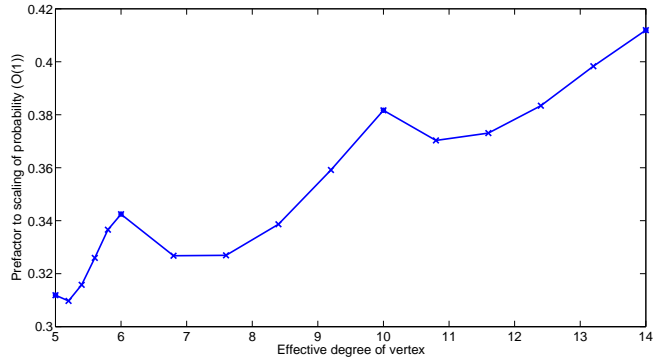}
\caption[Plot to show how the prefactor to the scaling of the maximum probability of the marked state varies with the connectivity in three dimensions.]{Plot to show how the prefactor obtained from the data in fig.~\ref{3dproball}, of the maximum probability of the marked state changes with the degree of the three dimensional structure being searched.}
\label{3dprobprefactor}
\end{minipage}
\end{figure}

The closer we can get this prefactor to unity, the lower the number of times we must run the algorithm. Figure \ref{3dprobprefactor} shows how this prefactor to the probability of finding the marked state varies with the degree of the structure being searched. We find the same `dips' and recurrences in the scaling as in the two dimensional case which can be explained in the same way.

%%%%%%%%%%%%%%%%%%%%%%%%
\section{Substrate disorder}
\label{sec:disorder}
%%%%%%%%%%%%%%%%%%%%%%%%

In the previous sections, we have only considered perfect, regular lattices with no defects. We now consider a simple form of noise (disorder) and are interested in how this affects the efficiency of the search algorithm. Previous work by Keating et al. \cite{keating07a} has highlighted the effect of Anderson localisation in continuous time quantum walks and also Krovi and Brun \cite{krovi05a,krovi06a,krovi07a} have shown how defects and a lack of symmetry in discrete time quantum walks can have an impact on the spreading of the walk. Both these factors suggest that the search algorithm will fail as soon as any level of disorder is introduced into the lattice. However, in contrast to these results, the study of the transport properties of discrete time quantum walks on 1D and 2D percolation lattices has recently been presented by Leung et al. \cite{leung10a}. They show that the spreading of a discrete time quantum walk, on a 2D percolation lattice, follows a fractional scaling, i.e. $\langle r \rangle \propto T^{\alpha}$ where $\langle r \rangle$ is the spread of the quantum walk and $T$ is the number of timesteps. This seems to be in contradiction to the previous work in the continuous time context. In addition, Abal et al. \cite{abal09a} have investigated how the quantum walk search algorithm performs in the presence of decoherence, specifially phase errors in the coin operator. In this work, we assume that we have a quantum computer with error correction available, and as such are not interested in these errors. Instead, we are interested in any disorder that could be present in an imperfect data structure. We aim to establish how much, if any, disorder the search algorithm can tolerate or if it fails completely. In order to do this, we use percolation lattices to allow us to vary the level of disorder in the lattice. Due to the computational time required for averaging over many lattices, we only consider site percolated lattices in this work, though we expect a qualitatively similar behaviour in lattices with edge percolation.

\subsection{Percolation lattices}
\label{sec:Ch_Eight_perc}

A percolation lattice is a lattice, for example a 2D Cartesian lattice, which has vertices (site percolation) or edges (bond percolation) randomly missing. The probability, $p$, of a vertex or edge existing determines the amount of disorder present in the lattice. As the probability increases there reaches a point, $p_{c}$, where the structure changes from a set of smaller, unconnected pieces into one larger piece which is almost all connected. At probabilities $p \ge p_{c}$, there will, in general, be a path from one side of the lattice to the other. We note here that this is only the case for structures with dimension two or more. It clear that any one dimensional lattice must be fully connected in order for a path to exist from one side of the lattice to the other, i.e. $p_{c} = 1$. 
\begin{figure}[!bt]
\begin{minipage}{\columnwidth}
\centering
\includegraphics[width=0.35\textwidth]{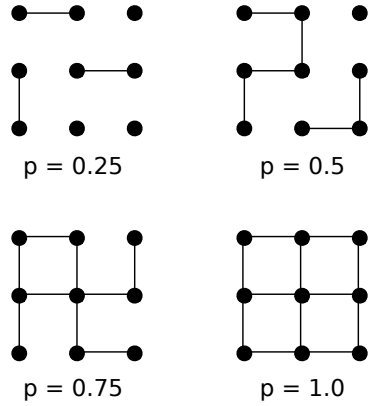}
\caption[An example of a 2D bond percolation lattice.]{An example of a 2D bond percolation lattice with varying levels of disorder determined by the probability of an edge existing. The critical percolation probability, $p_{c}=0.5$, for bond percolation clearly shows a path from one side of the lattice to the other.}
\label{perclattices}
\end{minipage}
\end{figure}
Figure \ref{perclattices} shows an example of a 2D bond percolation lattice with varying probability of an edge existing. A path from one side of the lattice can clearly be seen for probabilities greater than or equal to the critical percolation probability, $p_{c}=0.5$. This percolation threshold is only for bond percolation on a 2D square lattice. Although site and bond percolation lattices exhibit similar behaviour, the critical percolation probability differs, for site percolation $p_{c}=0.5928... .$ \cite{djordjevic82a,gebele84a}. Other lattices have varying critical probabilities depending on their structure, with many efficient numerical methods developed to calculate them \cite{djordjevic82a, gebele84a}. We are only interested here in two and three dimensional lattices, and we summarise the critical percolation probabilities of these in table \ref{percprobs}.
\begin{table}[hbt]
\caption{Summary of critical percolation probabilities for two and three dimensional lattices}
\centering
\begin{tabular}{c || c | c}
\hline \hline
Lattice & Bond & Site\\
\hline
2D & 0.5 & 0.5928..... \\
3D & 0.2488.... & 0.3116.....\\
\end{tabular}
\label{percprobs}
\end{table}

It is fairly obvious that at this critical percolation threshold, the properties of the lattice change significantly. For lattices with a percolation probability below the percolation threshold, it is clear that many of the sites in the lattice will be unreachable, whereas above the threshold the opposite is true (though perhaps through a less direct route than in a fully connected lattice). Due to their transport properties, percolation lattices are widely used to model various phenomena including forest fires, disease spread and the size and movement of oil deposits. For a good introduction to both the theory and use of percolation lattices, see Stauffer and Aharony \cite{stauffer92a}.

\subsection{Search algorithm on percolation lattices}
\label{sec:Ch_Eight_2dperc}

We are using the percolation lattices as a description for the database arrangement that we wish to run the quantum walk search algorithm upon. As the disorder introduced by using percolation lattices is random, we ran the search algorithm on many different percolation lattices (5000), and averaged over the results. It is obvious that at low probabilities of vertices (or edges) existing, that there may be sections of the graph that the quantum walk is unable to reach. In fact, at very low probabilities, it is likely that the marked state will be in a small, unconnected region of the lattice where it will never be `found'. In these cases, this means the marked state will only ever be able to attain a small portion of the total probability. We set the condition on the algorithm that the probability of the marked state must reach at least twice the value of the initial superposition in order for it to succeed. Similarly, the time to find this maximum probability is artificially smaller than it should be if the entire lattice was connected. This is due to the walker only having to coalesce on the marked state over a small piece of the lattice. In order to combat this, we set the time to find the marked state as zero if the algorithm failed. If it succeeded, we took the reciprocal of the time to find the marked state. After averaging over many different percolation lattices, we again took the reciprocal of this averaged time in order to give a clearer view on how the algorithm scaled with time. We also set the probability of the marked state to be zero if the algorithm failed. 

In order to run the quantum walk search algorithm on percolation lattices, we have to deal with the fact that the lattice is not $d$-regular. In this setting, we cannot just add self loops to make the lattice regular as in \cite{kempe03b} as we want to know exactly how the disorder affects the algorithm. Instead, we take the Grover coin for the degree of the vertex in question and `pad' it out with the identity operator for the edges that are missing. For example if we have a vertex with just edge 3 missing, the operator would be
\begin{equation}
G^{perc}_{1,2,4} = \left ( \begin{array}{cccc} -\frac{1}{3} & \frac{2}{3} & 0 & \frac{2}{3} \\ \frac{2}{3} & -\frac{1}{3} & 0 & \frac{2}{3}\\ 0 & 0 & 1 & 0 \\ \frac{2}{3} & \frac{2}{3} & 0 & -\frac{1}{3} \end{array} \right ),
\label{percgrover}
\end{equation}
where $G^{perc}_{1,2,4}$ represents the Grover coin with edges 1, 2 and 4 present. In the case of a two dimensional percolation lattice, there are 16 combinations of edges that can be present / missing. For a three dimensional percolation lattice, this increases to 64 combinations. In order to deal with this, we maintain the labelling of the edges as previously and assign a binary number to each edge, depending on whether an edge is present or not. The example above, eq.~(\ref{percgrover}), would therefore be $1101$. This creates the $2^{d}$ combinations  we require. There is then a fixed mapping between each binary number and the correct coin for each vertex. 

In addition to the coin operator changing, we must also modify the initial state to account for the missing vertices or edges. This could be done in several ways. We try to stick as closely to the initial state of the basic quantum walk search algorithm by just splitting the state into an equal superposition over all the possible edges present.

\subsection{Two dimensional percolation lattices}

\begin{figure}[!bt]
\begin{minipage}{\columnwidth}
\centering
\includegraphics[width=0.75\textwidth]{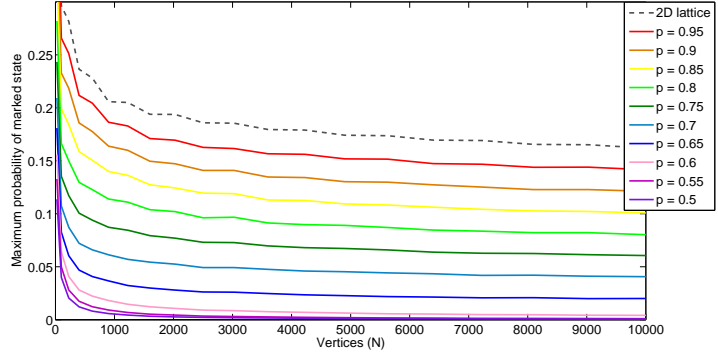}
\caption[Plot to show how the maximum probability of marked site varies with the size of the dataset and percolation probability for site percolation.]{Plot to show how the maximum probability of the marked state varies with the size of the dataset and percolation probability in two dimensions. We also show the same plot for a fully connected two dimension lattice (dashed line).}
\label{twodprobperc}
\end{minipage}
\end{figure}

We now show our initial results for the quantum walk search algorithm on two dimensional site percolation lattices. We firstly show, fig.~\ref{twodprobperc}, how the maximum probability of the marked state varies with both the size of the dataset and the percolation probability. We see, as we would think intuitively, that as the percolation probability drops and the structure becomes less connected, the maximum probability of the marked state decreases.
\begin{figure}[!bt]
\begin{minipage}{\columnwidth}
\centering
\includegraphics[width=0.75\textwidth]{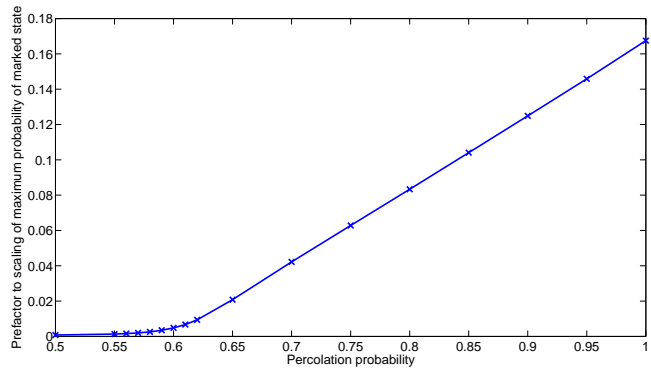}
\caption[Plot to show how the prefactor to the scaling of the maximum probability of the marked state varies with percolation probability for site percolation.]{Plot to show how the prefactor to the scaling of the maximum probability of the marked state, from the data in fig.~\ref{twodprobperc}, varies with the size of the dataset and the percolation probability for site percolation in two dimensions.}
\label{twodpercprobscaling}
\end{minipage}
\end{figure}

We note that the scaling of the maximum probability initially maintains the logarithmic scaling of the basic 2D lattice before eventually reverting to the scaling of the line, $1/N$, at lower percolation probabilities. In the case of site percolation, this change in scaling seems to occur at roughly probabilities below $p\approx0.65$, not significantly higher than the critical percolation threshold. This is expected as at the critical threshold, the structure has in general a single path from one side to the other, effectively a 1D lattice. Our numerical results match this behaviour, with the scaling of the probability of the marked state matching that of the line at this point. At percolation probabilities higher than the critical threshold, we see a change in the prefactor to the scaling of the maximum probability of the marked state. We show this prefactor to the logarithmic scaling in fig.~\ref{twodpercprobscaling}. It is easy to see that as soon as the percolation probability passes the critical threshold, $p_{c}=0.5928....$, the scaling increases in a linear fashion. We also note here, after investigation on a finer scale, that there is a gradual change in this prefactor scaling around the critical percolation threshold.

\begin{figure}[!tb]
\begin{minipage}{\columnwidth}
\centering
\includegraphics[width=0.75\textwidth]{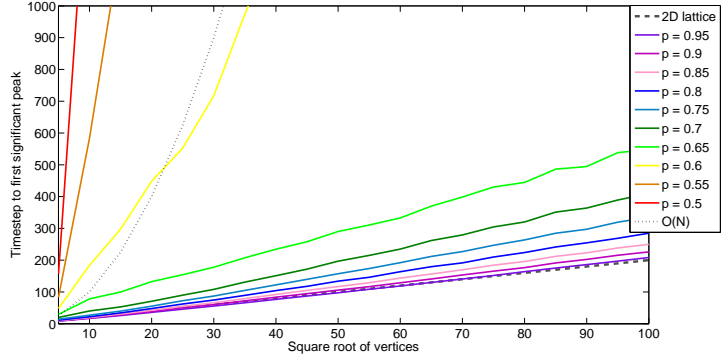}
\caption[Plot to show how the time to find the marked state varies with the size of the dataset and percolation probability for site percolation.]{Plot to show how the time to find the marked state varies with the size of the dataset and the percolation probability for site percolated lattices in two dimensions. We also show the same plot for a fully connected two dimensional lattice (dashed line).}
\label{twodtimesite}
\end{minipage}
\end{figure}

The time to find the marked state follows a similar behaviour, gradually changing from the quadratic scaling of the 2D lattice to a classical linear scaling as $p$ reduces. We show the time to find the marked state for site percolation in fig.~\ref{twodtimesite}. 
\begin{figure}[!tb]
\begin{minipage}{\columnwidth}
\centering
\includegraphics[width=0.75\textwidth]{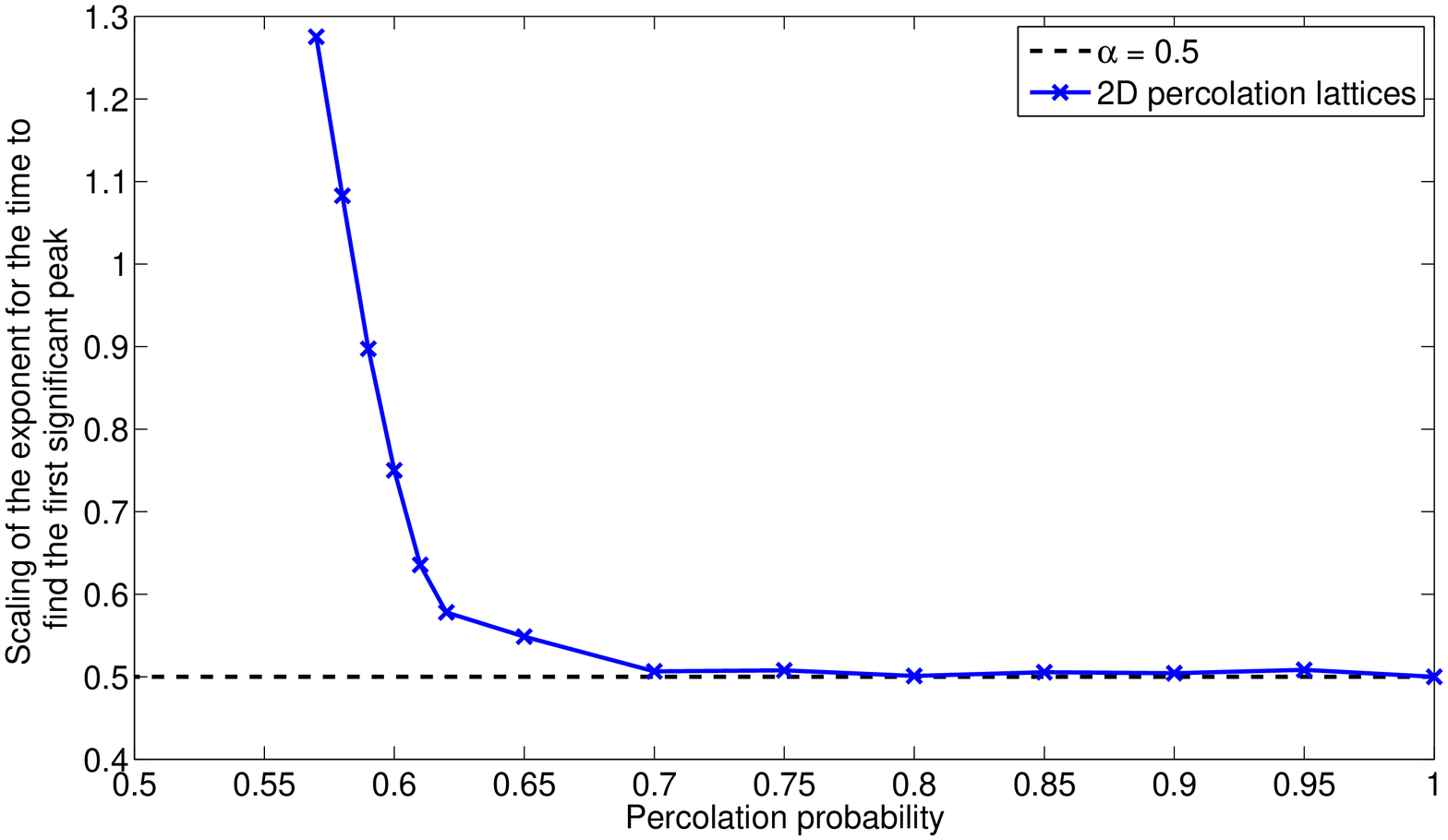}
\caption[Plot to show how the exponent, $\alpha$, to the scaling of the time to find the maximum probability of the marked state varies with percolation probability for site percolation.]{Plot to show how the exponent, $\alpha$, to the scaling of the time to find the maximum probability of the marked state, from the data in fig.~\ref{twodtimesite}, varies with the size of the dataset and the percolation probability for site percolation in two dimensions. Also shown is $\alpha=0.5$ to indicate the lower bound of the algorithm (dashed line).}
\label{twodperctimescaling}
\end{minipage}
\end{figure}
We see that when $p=0.6$, the scaling of the time to find the marked state is very similar to the classical run time, $O(N)$. The kinks in this scaling (and the other percolation probabilities) are just from averaging over many percolation lattices. Given more time, a higher number could be run and thus a smoother scaling obtained. It can be seen that the time to find the marked state seems to retain the quadratic quantum speed up, even in the presence of a non-trivial level of disorder. As in the work of Leung et al. \cite{leung10a}, it seems as though the scaling of the time to find the marked state may follow a fractional scaling from quadratic back to linear as,
\begin{equation}
T \propto N^{\alpha},
\end{equation}
where $T$ is the time to find the marked state and $N$ is the size of the dataset. We follow the analysis in \cite{leung10a} to establish how the scaling of the time to find the marked state varies with the percolation probability. We show, in fig.~\ref{twodperctimescaling}, how the value of the coefficient $\alpha$ varies as the level of disorder is increased. 

We can see the quadratic speedup is maintained, $\alpha\approx0.5$, for percolation probabilities of roughly $p>0.65$. Below this probability, the quantum speed up disappears gradually to end at the classical run time when $p=p_{c}$. This is for the same reason as in the scaling of the maximum probability of the marked state, at the critical threshold the structure is effectively a line. Below the critical threshold, the algorithm fails (the marked state is probably in a disconnected region). We note here that the coefficient, $\alpha$, is not exactly 0.5 as we expect for the quadratic speed up. This is most probably due to the fact that percolation lattices are random in nature, and we only average over a specific number. If we averaged over more, then we would see a more constant scaling of the coefficient at $\alpha=0.5$, i.e. a full quadratic speed up.  

\subsection{Three dimensional percolation lattices}

\begin{figure}[!bt]
\begin{minipage}{\columnwidth}
\centering
\includegraphics[width=0.75\textwidth]{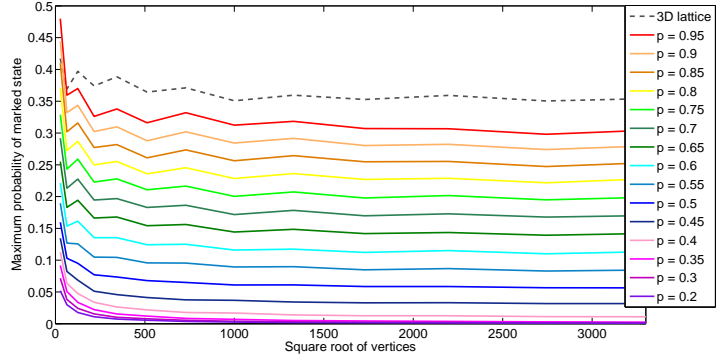}
\caption[Plot to show how the maximum probability of marked site varies with the size of the dataset and percolation probability for site percolation.]{Plot to show how the maximum probability of the marked state varies with the percolation probability in three dimensions. We also show the same plot for a fully connected three dimension lattice (dashed line).}
\label{threedprobperc}
\end{minipage}
\end{figure}

\begin{figure}[!bt]
\begin{minipage}{\columnwidth}
\centering
\includegraphics[width=0.75\textwidth]{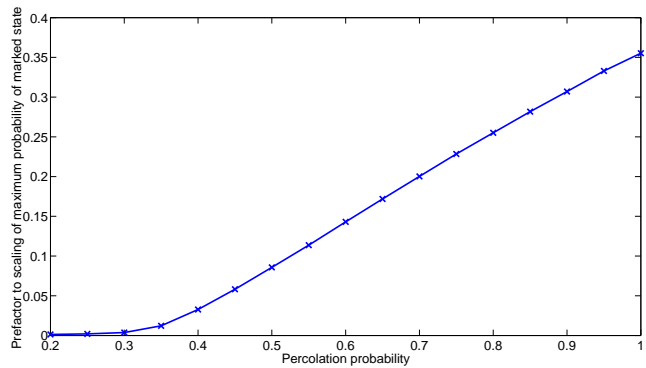}
\caption[Plot to show how the prefactor to the scaling of the maximum probability of the marked state varies with percolation probability for site percolation.]{Plot to show how the prefactor to the scaling of the maximum probability of the marked state, from the data in fig.~\ref{threedprobperc}, varies with the size of the dataset and the percolation probability for site percolation in three dimensions.}
\label{threedpercprobscaling}
\end{minipage}
\end{figure}

We now turn our attention to three dimensional site percolation lattices. We follow the same analysis as in the two dimensional case. We firstly show, fig.~\ref{threedprobperc}, how the maximum probability of the marked state varies as the percolation probability is decreased. We see, as in the two dimensional case, that the basic scaling of the maximum probability matches that of the three dimensional lattice until the percolation probability drops to roughly the critical percolation threshold, $p_{c} = 0.3116...$. 
We show in fig.~\ref{threedpercprobscaling}, how the prefactor to this scaling of the maximum probability varies with the percolation probability. In the same way as the two dimensional case, we see an almost linear scaling of the prefactor once the percolation probability has passed the critical threshold. The scaling here doesn't seem to be as close as in the two dimensional case. This is probably because in the case of three dimensional percolation lattices, there are many more combinations of lattice which can be created. Averaging over more of these lattices would most probably give a smoother fit.

\begin{figure}[!bt]
\begin{minipage}{\columnwidth}
\centering
\includegraphics[width=0.75\textwidth]{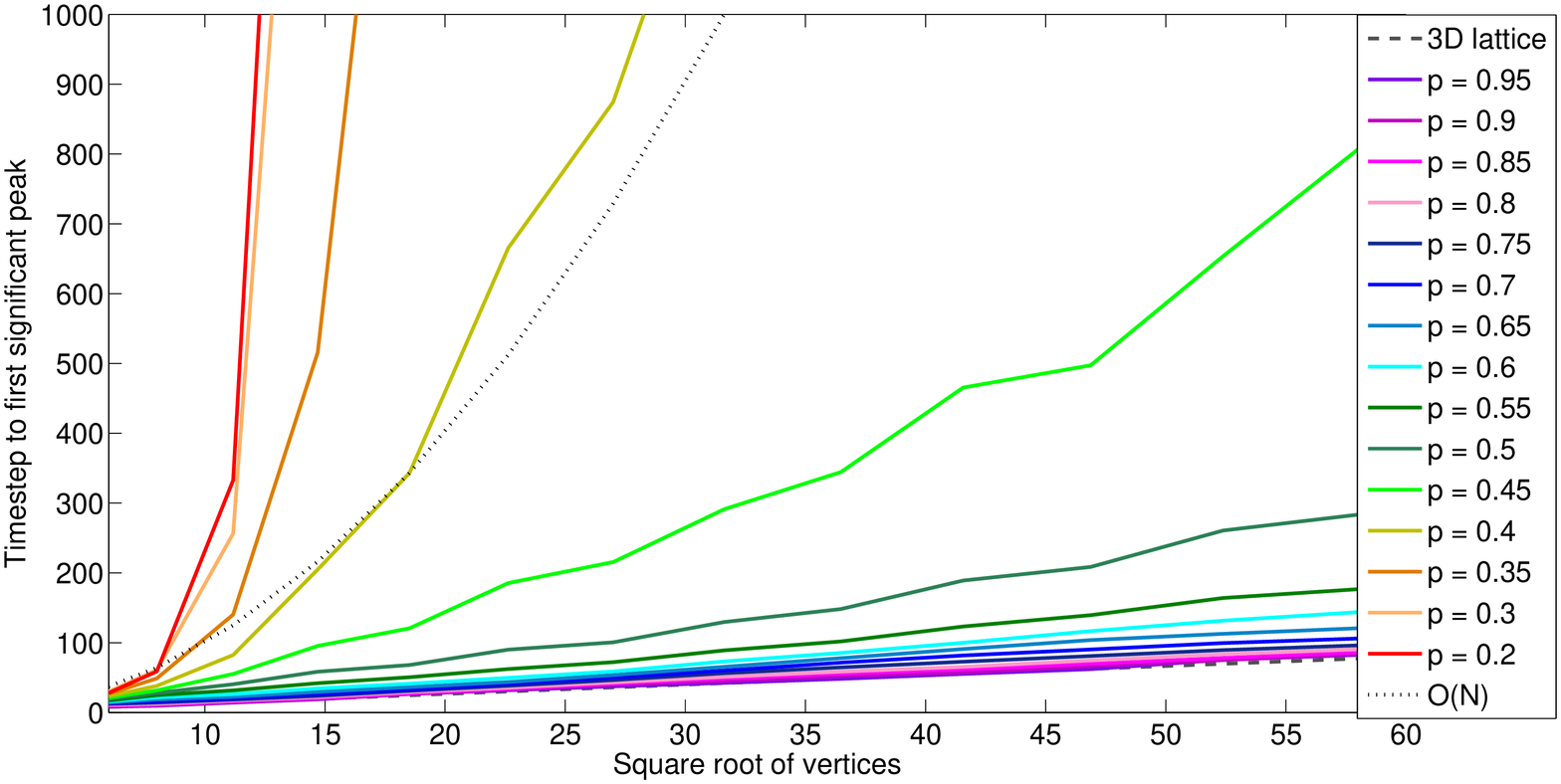}
\caption[Plot to show how the time to find the marked state varies with the size of the dataset and percolation probability for site percolation.]{Plot to show how the time to find the marked state varies with the size of the dataset and the percolation probability for site percolation lattices in three dimensions. We also show the same plot for a fully connected three dimension lattice (dashed line).}
\label{threedtimesite}
\end{minipage}
\end{figure}

\begin{figure}[!bt]
\begin{minipage}{\columnwidth}
\centering
\includegraphics[width=0.75\textwidth]{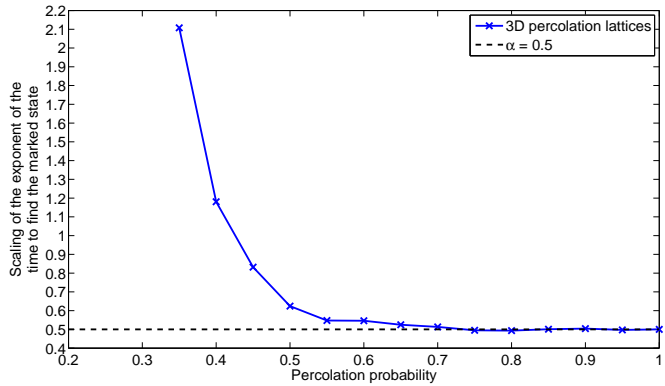}
\caption[Plot to show how the exponent, $\alpha$, to the scaling of the time to find the maximum probability of the marked state varies with percolation probability for site percolation.]{Plot to show how the exponent, $\alpha$, to the scaling of the time to find the maximum probability of the marked state, from the data in fig.~\ref{threedtimesite}, varies with the size of the dataset and the percolation probability for site percolation in three dimensions. Also shown is $\alpha=0.5$ to indicate the lower bound of the algorithm (dashed line).}
\label{threedperctimescaling}
\end{minipage}
\end{figure}

The time to find the marked state, in the three dimensional case, follows the same behaviour as in the two dimensional percolation lattices. We show in fig.~\ref{threedtimesite}, how the time to find the marked state varies with the percolation probability. We see, fig.~\ref{threedperctimescaling}, as in the two dimensional case, that the scaling coefficient, $\alpha$, gradually changes from the quadratic speed up to the classical run time. Again, we note that the quadratic speed up is maintained for a non-trivial amount of disorder before gradually changing to the classical run time at the point $p=p_{c}$. 
We do note, as in the two dimensional case, that the coefficient is not exactly 0.5. This can be explained in the same way as the two dimensional percolation lattices, and averaging over more lattices should give a constant value of the coefficient $\alpha$.

%%%%%%%%%%%%%%%%%%%%%%%%%%%%%%%%%%%%%%%%%%%%
\section{Discussion}
\label{sec:discussion}
%%%%%%%%%%%%%%%%%%%%%%%%%%%%%%%%%%%%%%%%%%%%

In this paper, we have discussed various factors which affect the efficiency of the quantum walk search algorithm. We introduce a simple form of tunnelling which allows us to modify the substrate we use as the database arrangement, and use this to interpolate between structures with varying dimensionality and degree. We find that although the dependence on the spatial dimension of the underlying substrate is strong, it is not the only factor which affects the efficiency of the algorithm. We also find secondary dependencies on the connectivity and symmetry of the structure. In addition, we use percolation lattices to model disorder in the lattice in a simple way. In this case we find, counter-intuitively, that the algorithm is able to maintain the quantum speed up even in the presence of non trivial levels of disorder.  We now discuss our findings for each factor in turn.

\subsection{Dimensional dependence}

We have shown two different ways in which we can interpolate between structures of differing spatial dimension. Firstly, we use a tunnelling operator to vary specific edges of a lattice enabling us to gradually change the spatial dimension of the lattice. In this case, we find a sudden change in the scaling of the maximum probability of the marked state as soon as there is even a very small probability of the edges existing. This seems to indicate that the `strength' of the edges in the lattice is of little importance, with the dependence on the specific spatial dimension taking precedence. However, we find that the prefactor to the scaling of this probability varies with the strength of the tunnelling edges, increasing as the tunnelling strength increases. The basic scaling of the time to find the marked state is not affected by the change in dimensionality, we note though that the prefactor to the scaling decreases as the tunnelling strength increases, hence the algorithm becomes more efficient.

The other case we consider is the case of lattices with varying height or depth, for example, a 3D lattice with fixed width and height but of varying depth. Although this structure is still strictly three dimensional, when the depth is very low and the width (height) is large, the quantum walker will see the structure as almost a basic 2D Cartesian lattice. Suprisingly, in this case we see a gradual change in scaling in the maximum probability of the marked state. At low depths of the lattice, the scaling is almost the same as the lower spatial dimensional structure gradually changing to the higher dimensional structure scaling as the depth increases to become equal to that of the other dimensions. This highlights the importance of full symmetry in the quantum walk search algorithm.

\subsection{Connectivity}

We show how the search algorithm is affected by varying connectivity in regular lattices. We use our simple model of tunnelling to allow us to interpolate between structures such as the square lattice ($d=4$) and the triangular lattice ($d=6$). With this model, we are able to identify how the prefactors to the scaling of both the maximum probability of the marked state and the time to find the marked state vary with the connectivity of the structure.

The basic scaling of the time to find the marked state, $O(\sqrt{N})$, is not affected by the increase in connectivity but we find the prefactor to this scaling reduces as the connectivity of the structure being searched increases. This is due to the additional paths the walker can take to coalesce on the marked state, thus increasing the efficiency of the algorithm in both two and three dimensions. 

The maximum probability of the marked state is also affected by the connectivity of the underlying structure. We find that the additional connectivity does not affect the basic scaling of $O(1/\log_{2} N)$ in the two dimensional case. Only moving to three spatial dimensions allows the walker to find the marked state with a constant probability, $O(1)$. However, we do note that in both two and three dimensions the prefactors to this scaling, in general, increase as the connectivity of the structure increases. Again, this increases the efficiency of the algorithm as it may not have to be repeated so many times. We also find that the probability of the marked state does not increase uniformly with the additional connectivity. We see the prefactor in the scaling drop and then recover itself before increasing as the tunnelling strength increases. This is due to the dynamics of the quantum walk on a structure with some broken symmetry, i.e. low tunnelling strength between vertices. We briefly investigated the dynamics of the walk by starting the walker in a single location and monitoring how quickly it spread outwards with varying tunnelling strengths. This confirmed our results for the search algorithm as we found that the spread of the quantum walk also dropped for lower tunnelling strengths before recovering and eventually increasing at higher tunnelling probabilities. However, this work on the spreading of the walk compared to tunnelling strength is by no means exhaustive and it would be interesting to look more deeply into this in the future.

\subsection{Substrate disorder}

We studied both two and three dimensional percolation lattices as a way to model disorder in the quantum walk search algorithm. We are interested in how the algorithm performs with increasing disorder. We use percolation lattices as a random substrate for the database arrangement we wish to search. 

We find, in both the two and three dimensional cases, that as the level of disorder increases, the maximum probability of the marked state decreases. Whilst the percolation probability is higher than the critical percolation threshold, the basic scaling of the maximum probability of the marked state matches that of the basic lattice (in that spatial dimension). Once the percolation probability drop to the critical threshold, this scaling changes to that of the line, $1/N$. This is expected as at this point the structure is effectively a line. We also note the prefactor to the scaling of the maximum probability of the marked state increases linearly once the percolation probability is greater than the critical threshold. 

The time to find the marked state follows a similar behaviour. We find that as the disorder increases, the time to find the marked state also increases. Surprisingly though, we note that the quadratic speed up is maintained for a non-trivial level of disorder, before gradually reverting to the classical run time, $O(N)$, as the disorder reaches the critical percolation threshold. This seems to match the results of \cite{leung10a}, which show a fractional scaling for the spreading of the quantum walk from a maximal quantum spreading to a classical spreading at and below the critical threshold. However, this is in contrast to the work of Krovi and Brun \cite{krovi05a,krovi06a,krovi07a} who highlight the effect of localisation on the quantum walk when defects are introduced into the substrate.

Both these factors indicate that the quantum walk search algorithm seems to be more robust to the effects of disorder and symmetry than the basic spreading of the quantum walk. This could be due to the fact that the initial state of the walker is spread across the whole lattice. We have seen that the algorithm becomes less efficient as the disorder increases, but at percolation probabilities greater than the critical threshold, the algorithm still seems to be viable, although more amplification of the result may be required.  
\\
\\
\textit{Acknowledgments}
The authors would like to thank Jiannis Pachos and Noah Linden for useful and interesting discussions.
NL was funded by the UK Engineering and Physical Sciences Research Council. VK is funded by a UK Royal Society University Research Fellowship.


\begin{thebibliography}{99}

\bibitem[Aaronson and Ambainis 2003]{aaronson03a}
S.~Aaronson and A.~Ambainis.
\newblock {Quantum search of spatial regions}.
\newblock In {\em Proceedings of the 44th annual IEEE symposium on foundations
  of computer science (FOCS), 2003}, pages 200--209. IEEE, 2003.

\bibitem[Abal et al. 2009]{abal09a}
G.~Abal, R.~Donangelo, F.~L. Marquezino, A.~C. Oliveira, and R.~Portugal.
\newblock {Decoherence in Search Algorithms}.
\newblock {\em Proceedings of the 29th Brazilian computer society congress
  (SEMISH)}, pages 293--306, 2009.

\bibitem[Abal et al. 2010]{abal10a}
G.~Abal, R.~Donangelo, F.~L. Marquezino, and R.~Portugal.
\newblock {Spatial search in a honeycomb network}.
\newblock {\em To appear in Mathematical Structures in Computer Science. Arxiv
  preprint arXiv:1001.1139}, 2010.

\bibitem[Aharonov et al. 2001]{aharonov00a}
D.~Aharonov, A.~Ambainis, J.~Kempe, and U.~Vazirani.
\newblock {Quantum walks on graphs}.
\newblock In {\em Proceedings of the 33rd annual ACM symposium on theory of
  computing (STOC)}, pages 50--59. ACM, 2001.

\bibitem[Ambainis 2003]{ambainis04b}
A.~Ambainis.
\newblock {Quantum walks and their algorithmic applications}.
\newblock {\em International Journal of Quantum Information}, 1:507--518, 2003.

\bibitem[Ambainis 2004]{ambainis04c}
A.~Ambainis.
\newblock {Quantum walk algorithm for element distinctness}.
\newblock In {\em Proceedings of the 45th annual IEEE symposium on foundations
  of computer science (FOCS), 2004}, pages 22--31. IEEE, 2004.

\bibitem[Ambainis et al. 2001]{ambainis01a}
A.~Ambainis, E.~Bach, A.~Nayak, A.~Vishwanath, and J.~Watrous.
\newblock {One-dimensional quantum walks}.
\newblock In {\em Proceedings of the 33rd annual ACM symposium on theory of
  computing (STOC)}, pages 37--49. ACM, 2001.

\bibitem[Ambainis Kempe and Rivosh 2005]{ambainis04a}
A.~Ambainis, J.~Kempe, and A.~Rivosh.
\newblock {Coins make quantum walks faster}.
\newblock In {\em Proceedings of the 16th annual ACM-SIAM symposium on discrete
  algorithms (SODA)}, pages 1099--1108. Society for Industrial and Applied
  Mathematics, 2005.

\bibitem[Benioff 2002]{benioff00a}
P.~Benioff.
\newblock {Space searches with a quantum robot}.
\newblock {\em Quantum Computation and Information}, 305:1, 2002.

\bibitem[Bernstein et al. 1997]{bennett97a}
E.~Bernstein, C.~Bennett, G.~Brassard, and U.~Vazirani.
\newblock {Strengths and weaknesses of quantum computing}.
\newblock {\em SIAM Journal of Computing}, 26:1510--1523, 1997.

\bibitem[Brassard 2002]{brassard02a}
G.~Brassard, P.~H{\o}yer, M.~Mosca, and A.~Tapp.
\newblock {Quantum amplitude amplification and estimation}.
\newblock In {\em Quantum Computation and Information}, volume 305, pages
  53--74. American Mathematical Society, Providence, RI., 2002.

\bibitem[Buhrman and Spalek 2004]{buhrman04a}
H.~Buhrman and R.~Spalek.
\newblock {Quantum verification of matrix products}.
\newblock {\em Proceedings of the 17th annual ACM-SIAM symposium on discrete
  algorithms, 2004}, 2004.

\bibitem[Childs 2009]{childs09a}
A.~M. Childs.
\newblock {Universal computation by quantum walk}.
\newblock {\em Physical Review Letters}, 102:180501, 2009.

\bibitem[Childs and Goldstone 2004]{childs04a}
A.~M. Childs and J.~Goldstone.
\newblock {Spatial search and the Dirac equation}.
\newblock {\em Physical Review A}, 70:42312, 2004.

\bibitem[Childs and Goldstone 2004]{childs03a}
A.~M. Childs and J.~Goldstone.
\newblock {Spatial search by quantum walk}.
\newblock {\em Physical Review A}, 70:22314, 2004.

\bibitem[Djordjevic 1982]{djordjevic82a}
Z.~V. Djordjevic, H.~E. Stanley, and A.~Margolina.
\newblock {Site percolation threshold for honeycomb and square lattices}.
\newblock {\em Journal of Physics A: Mathematical and General}, 15:L405, 1982.

\bibitem[Farhi and Gutmann 1998]{farhi98a}
E.~Farhi and S.~Gutmann.
\newblock {Quantum computation and decision trees}.
\newblock {\em Physical Review A}, 58:915--928, 1998.

\bibitem[Gebele 1984]{gebele84a}
T.~Gebele.
\newblock {Site percolation threshold for square lattice}.
\newblock {\em Journal of Physics A: Mathematical and General}, 17:L51, 1984.

\bibitem[Gottlieb 2005]{gottlieb05a}
A.~D. Gottlieb.
\newblock {Convergence of continuous-time quantum walks on the line}.
\newblock {\em Physical Review E}, 72:47102, 2005.

\bibitem[Grimmett Janson and Scudo 2004]{grimmett04a}
G.~Grimmett, S.~Janson, and P.~F. Scudo.
\newblock {Weak limits for quantum random walks}.
\newblock {\em Physical Review E}, 69:26119, 2004.

\bibitem[Grover 1996]{grover96a}
L.~K. Grover.
\newblock {A fast quantum mechanical algorithm for database search}.
\newblock {\em Proceedings of the 28th annual ACM symposium on theory of
  computing (STOC), 1996}, page 212, 1996.

\bibitem[Hein and Tanner 2009]{hein09a}
B.~Hein and G.~Tanner.
\newblock {Quantum search algorithms on the hypercube}.
\newblock {\em Journal of Physics A: Mathematical and Theoretical}, 42:085303,
  2009.

\bibitem[Hein and Tanner 2010]{hein10a}
B.~Hein and G.~Tanner.
\newblock {Quantum search algorithms on a regular lattice}.
\newblock {\em Physical Review A}, 82:12326, 2010.

\bibitem[Inui Konishi and Konno 2004]{inui04a}
N.~Inui, Y.~Konishi, and N.~Konno.
\newblock Localization of two-dimensional quantum walks.
\newblock {\em Physical Review A}, 69(5):052323, 2004.

\bibitem[Keating et al. 2007]{keating07a}
J.~P. Keating, N.~Linden, J.~C.~F Matthews, and A.~Winter.
\newblock {Localization and its consequences for quantum walk algorithms and
  quantum communication}.
\newblock {\em Physical Review A}, 76:12315, 2007.

\bibitem[Kempe 2003]{kempe03b}
J.~Kempe.
\newblock {Quantum random walks: an introductory overview}.
\newblock {\em Contemporary Physics}, 44:307--327, 2003.

\bibitem[Kempe 2003]{kempe03a}
J.~Kempe.
\newblock {Quantum random walks hit exponentially faster}.
\newblock In {\em Lecture Notes in Computer Science - Proceedings of
  RANDOM'03}, volume 2764, pages 354--369. Springer, 2003.

\bibitem[Kendon 2003]{kendon03b}
V.~Kendon.
\newblock {Quantum walks on general graphs}.
\newblock {\em International Journal of Quantum Information}, 4:791, 2003.

\bibitem[Kendon 2007]{kendon06a}
V.~Kendon.
\newblock {Decoherence in quantum walks - a review}.
\newblock {\em Mathematical Structures in Computer Science}, 17:1169--1220,
  2007.

\bibitem[Kendon and Tregenna 2003]{kendon03c}
V.~Kendon and B.~Tregenna.
\newblock {Decoherence can be useful in quantum walks}.
\newblock {\em Physical Review A}, 67:42315, 2003.

\bibitem[Krovi and Brun 2006]{krovi05a}
H.~Krovi and T.~A. Brun.
\newblock {Hitting time for quantum walks on the hypercube}.
\newblock {\em Physical Review A}, 73:32341, 2006.

\bibitem[Krovi and Brun 2006]{krovi06a}
H.~Krovi and T.~A. Brun.
\newblock {Quantum walks with infinite hitting times}.
\newblock {\em Physical Review A}, 74:42334, 2006.

\bibitem[Krovi and Brun 2007]{krovi07a}
H.~Krovi and T.~A. Brun.
\newblock {Quantum walks on quotient graphs}.
\newblock {\em Physical Review A}, 75:62332, 2007.

\bibitem[Krovi et al. 2010]{krovi10a}
H.~Krovi, F.~Magniez, M.~Ozols, and J.~Roland.
\newblock {Finding is as easy as detecting for quantum walks}.
\newblock {\em Automata, Languages and Programming}, pages 540--551, 2010.

\bibitem[Leung et al. 2010]{leung10a}
G.~Leung, P.~Knott, J.~Bailey, and V.~Kendon.
\newblock {Coined quantum walks on percolation graphs}.
\newblock {\em New Journal of Physics}, 12:123018, 2010.

\bibitem[Lovett et al. 2010]{lovett10a}
N.~B. Lovett, S.~Cooper, M.~Everitt, M.~Trevers, and V.~Kendon.
\newblock {Universal quantum computation using the discrete-time quantum walk}.
\newblock {\em Physical Review A}, 81:42330, 2010.

\bibitem[Lovett et al. 2011]{lovett10b}
Neil Lovett, Matthew Everitt, Matthew Trevers, Daniel Mosby, Dan Stockton, and
  Viv Kendon.
\newblock Spatial search using the discrete time quantum walk.
\newblock {\em Natural Computing}, pages 1--13, 2011.
\newblock 10.1007/s11047-011-9279-4.

\bibitem[Mackay et al. 2002]{mackay02a}
T.~D. Mackay, S.~D. Bartlett, L.~T. Stephenson, and B.~C. Sanders.
\newblock {Quantum walks in higher dimensions}.
\newblock {\em Journal of Physics A: Mathematical and General}, 35:2745, 2002.

\bibitem[Magniez and Nayak 2005]{magniez05b}
F.~Magniez and A.~Nayak.
\newblock {Quantum complexity of testing group commutativity}.
\newblock {\em Automata, Languages and Programming}, pages 1312--1324, 2005.

\bibitem[Magniez et al. 2009]{magniez08a}
F.~Magniez, A.~Nayak, P.~C. Richter, and M.~Santha.
\newblock {On the hitting times of quantum versus random walks}.
\newblock In {\em Proceedings of the 20th annual ACM-SIAM symposium on discrete
  algorithms (SODA)}, pages 86--95. Society for Industrial and Applied
  Mathematics, 2009.

\bibitem[Magniez et al. 2007]{magniez07a}
F.~Magniez, A.~Nayak, J.~Roland, and M.~Santha.
\newblock {Search via quantum walk}.
\newblock In {\em Proceedings of the 39th annual ACM symposium on theory of
  computing (STOC)}, pages 575--584. ACM, 2007.

\bibitem[Magniez Santha and Szegedy 2005]{magniez05a}
F.~Magniez, M.~Santha, and M.~Szegedy.
\newblock {Quantum algorithms for the triangle problem}.
\newblock In {\em Proceedings of the 16th annual ACM-SIAM symposium on discrete
  algorithms (SODA)}, pages 1109--1117. Society for Industrial and Applied
  Mathematics, 2005.

\bibitem[Marquezino Portugal and Abal 2010]{marquezino10a}
F.~L. Marquezino, R.~Portugal, and G.~Abal.
\newblock {Mixing Times in Quantum Walks on Two-Dimensional Grids}.
\newblock {\em Physical Review A}, 82:042341, 2010.

\bibitem[Moore and Russell 2002]{moore01a}
C.~Moore and A.~Russell.
\newblock {Quantum walks on the hypercube}.
\newblock {\em Proceedings of 6th international workshop on randomization and
  approximation techniques in computer science (RANDOM)}, pages 164--178, 2002.

\bibitem[Nayak and Vishwanath 2000]{nayak00a}
A.~Nayak and A.~Vishwanath.
\newblock {Quantum walk on the line}.
\newblock {\em DIMACS Technical Report 2000-43, Arxiv preprint arXiv:0010117},
  2000.

\bibitem[Patel et al. 2010]{patel10b}
A.~Patel, K.~S. Raghunathan, and Md.~A. Rahaman.
\newblock {Search on a Hypercubic Lattice through a Quantum Random Walk: II.
  d=2}.
\newblock {\em Physical Review A}, 82:032331, 2010.

\bibitem[Patel and Rahaman 2010]{patel10a}
A.~Patel and Md.~A. Rahaman.
\newblock {Search on a Hypercubic Lattice through a Quantum Random Walk: I.
  d>2}.
\newblock {\em Physical Review A}, 82:032330, 2010.

\bibitem[Poto{\v{c}}ek et al. 2009]{potocek08a}
V.~Poto{\v{c}}ek, A.~G{\'a}bris, T.~Kiss, and I.~Jex.
\newblock {Optimized quantum random-walk search algorithms on the hypercube}.
\newblock {\em Physical Review A}, 79:12325, 2009.

\bibitem[Reitzner et al. 2009]{reitzner09a}
D.~Reitzner, M.~Hillery, E.~Feldman, and V.~Bu{\v{z}}ek.
\newblock {Quantum searches on highly symmetric graphs}.
\newblock {\em Physical Review A}, 79:12323, 2009.

\bibitem[Santha 2008]{santha08a}
M.~Santha.
\newblock {\em {Quantum walk based search algorithms}}.
\newblock Theory and Applications of Models of Computation. Springer, 2008.
\newblock 31--46.

\bibitem[Santos and Portugal 2009]{santos09a}
R.~A.~M. Santos and R.~Portugal.
\newblock {Quantum Hitting Time on the Complete Graph}.
\newblock {\em To appear in International Journal of Quantum Information. Arxiv
  preprint arXiv:0912.1217}, 2009.

\bibitem[Shenvi Kempe and Whaley 2003]{shenvi02a}
N.~Shenvi, J.~Kempe, and K.~B. Whaley.
\newblock {A quantum random-walk search algorithm}.
\newblock {\em Physical Review A}, 67:52307, 2003.

\bibitem[Stauffer and Aharony 1992]{stauffer92a}
D.~Stauffer and A.~Aharony.
\newblock {\em {Introduction to percolation theory}}.
\newblock Taylor \& Francis, 1992.

\bibitem[Szegedy 2004]{szegedy04a}
M.~Szegedy.
\newblock {Quantum Speed-Up of Markov Chain Based Algorithms}.
\newblock In {\em Proceedings of 45th annual IEEE symposium on foundations of
  computer science (FOCS)}, pages 32--41. IEEE, 2004.

\bibitem[Tregenna et al. 2003]{kendon03a}
B.~Tregenna, W.~Flanagan, R.~Maile, and V.~Kendon.
\newblock {Controlling discrete quantum walks: coins and initial states}.
\newblock {\em New Journal of Physics}, 5:83, 2003.

\bibitem[Tulsi 2008]{tulsi08a}
A.~Tulsi.
\newblock {Faster quantum-walk algorithm for the two-dimensional spatial
  search}.
\newblock {\em Physical Review A}, 78:12310, 2008.

\bibitem[Underwood and Feder 2010]{underwood10a}
M.~S. Underwood and D.~L. Feder.
\newblock {Universal quantum computation by discontinuous quantum walk}.
\newblock {\em Physical Review A}, 82:042304, 2010.

\bibitem[Zalka 1999]{zalka99a}
C.~Zalka.
\newblock {Grover$'$s quantum searching algorithm is optimal}.
\newblock {\em Physical Review A}, 60:2746--2751, 1999.

\end{thebibliography}
\end{document}